\documentclass[aps,prl,reprint,nofootinbib]{revtex4-1}
\usepackage{graphics,color,graphicx,amsmath}
\usepackage{xr}
\usepackage{multibib}
\newcites{app}{Appendix}

\newcommand{\br}[1]{\left<#1\right>}

\begin{document}
\title{Emergent regularities and scaling in armed conflict data}
\author{Edward D. Lee,$^{1,2}$ Bryan C. Daniels,$^{3}$ Christopher R. Myers,$^{1,4}$ David C. Krakauer,$^2$ Jessica C. Flack$^2$}
\affiliation{$^1$Department of Physics, 142 Sciences Dr, Cornell University, Ithaca NY 14853\\
$^{2}$Santa Fe Institute, 1399 Hyde Park Rd, Santa Fe, NM 87501\\
$^{3}$ASU--SFI Center for Biosocial Complex Systems, Arizona State University, Tempe, AZ 85287\\
$^{4}$Center for Advanced Computing, Cornell University, Ithaca, NY 14853
}

\date{\today}

\begin{abstract}
Armed conflict exhibits regularities beyond known power law distributions of fatalities and duration over varying culture and geography. We systematically cluster conflict reports from a database of $10^5$ events from Africa spanning 20 years into {\it conflict avalanches}. Conflict profiles collapse over a range of scales. Duration, diameter, extent, fatalities, and report totals satisfy mutually consistent scaling relations captured with a model combining geographic spread and local conflict-site growth. The emergence of such social scaling laws hints at principles guiding conflict evolution.
\end{abstract}

\maketitle

In the 1940s, Richardson famously noted the distribution of fatalities in warfare followed a power law \cite{richardsonFrequencyOccurrence1941}. Since then power law statistics in armed conflict have been observed across a variety of data sets including terrorism and conventional warfare \cite{clausetFrequencySevere2007,picoliUniversalBursty2015,gillespieEstimatingNumber2017,clausetTrendsFluctuations2018}. Though mechanisms underlying these regularities remain elusive, possibilities have been discussed in the context of cellular-automata \cite{cedermanModelingSize2003}, coalescence-fragmentation \cite{bohorquezCommonEcology2009}, and self-organized critical forest fire models \cite{robertsFractalitySelfOrganized1998}. Observation of self-similarity in conflict statistics has inspired discussion of criticality as is described by renormalization group theory, a framework for organizing many microscopic mechanisms into universality classes distinguished by their macroscopic properties \cite{cardyScalingRenormalization1996,Binney:1993wu}. Renormalization group analysis of nonequilibrium critical phenomena explains why at large scales a low-dimensional description emerges, leading to a self-consistent scaling framework with a rich array of predictions including consistent scaling relations and universal scaling functions \cite{sethnaCracklingNoise2001}. Building on this perspective, we propose a scaling framework that unifies multiple properties of conflict with one another. Then, we develop a random, branching model of conflict that includes local conflict growth, dissipation, and regional variation in intensity to show armed conflicts are dominated by a low-dimensional process that scales with physical dimensions in a surprisingly unified and predictable way.


\begin{figure}[b]
	\includegraphics[width=.9\linewidth,clip=true,trim=0 5 0 5]{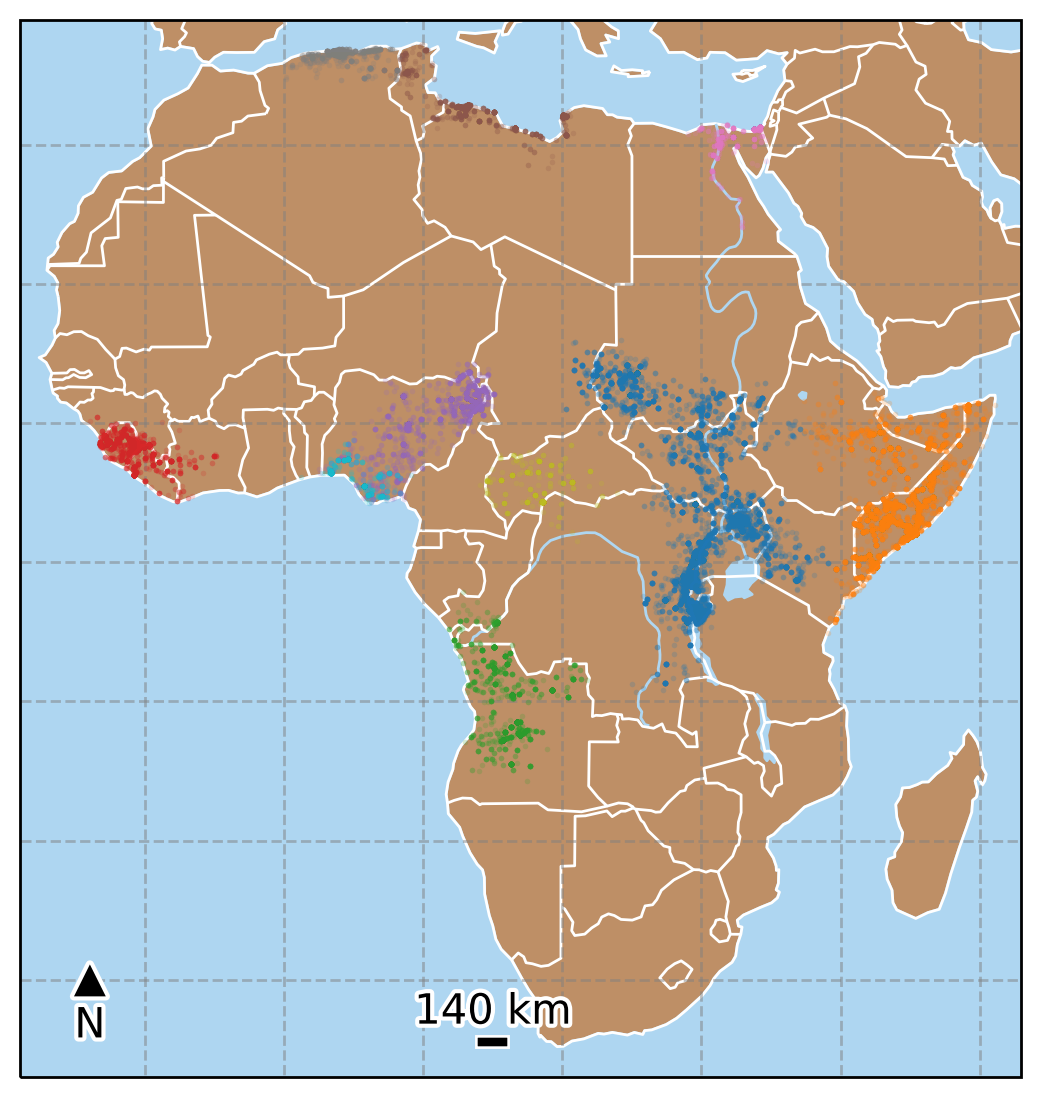}
	\caption{Battle avalanches in Africa between 1997--2016 \cite{raleighIntroducingACLED2010}. Spatial distribution of 10 largest conflict avalanches by reports $R$ for given separation scales $b=140\,{\rm km}$ and $a=128\,{\rm days}$. Spatial density is highly non-uniform, largely confined to land, and typically denser near population centers.}\label{gr:fig0}
\end{figure}

\begin{figure}[tb]\centering
	\includegraphics[width=\linewidth]{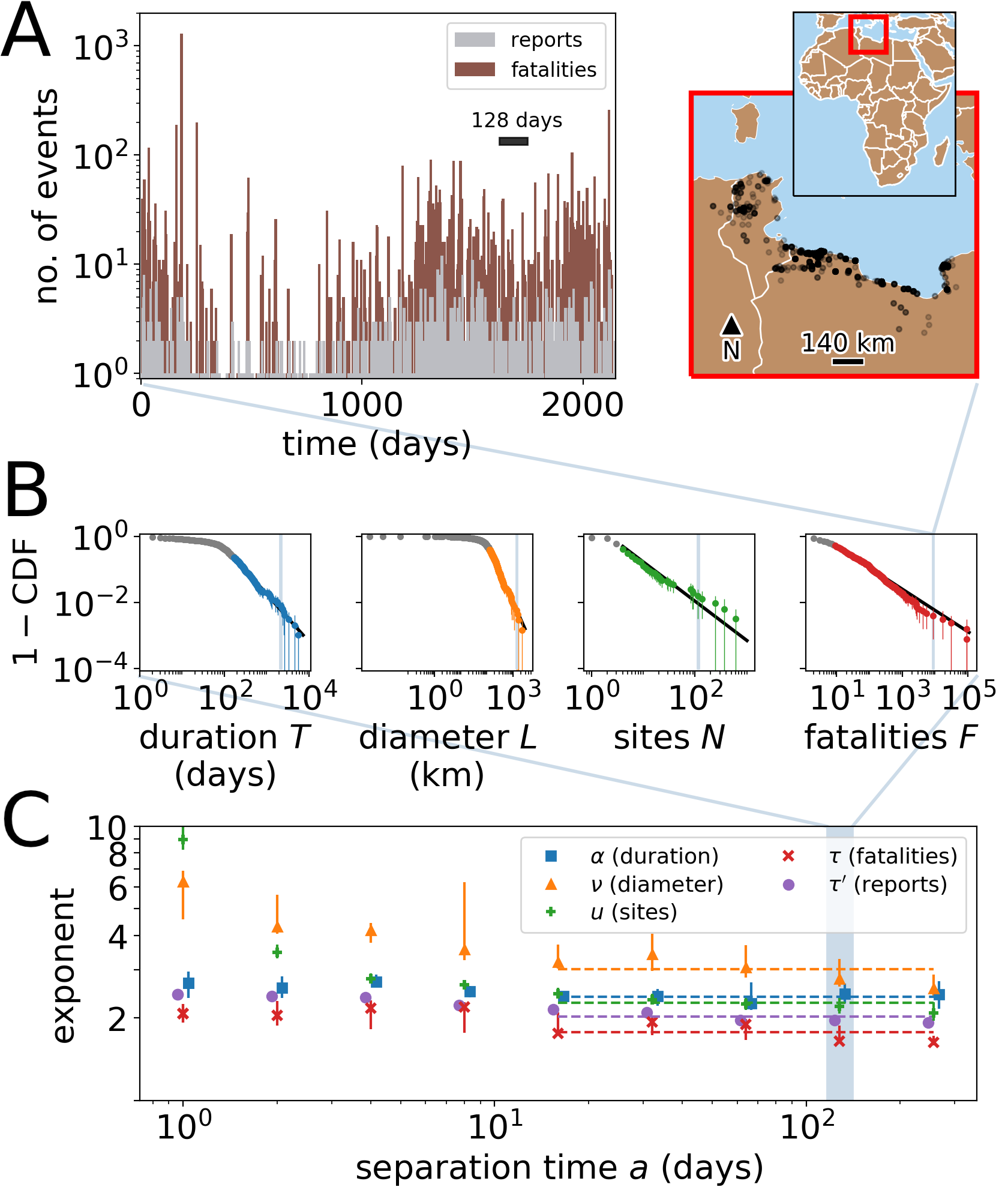}
\caption{(A) A single conflict avalanche erupting across Tunisia and Libya from Feb. 2, 2011 til Dec. 27, 2016 with temporal profile on left and spatial distribution on right. This avalanche lasts $T=2{,}141$\,days, extends $L=1{,}364$\,km, covers $N=95$ geographic sites of radius 70\,km, and has $F=8{,}569$ reported fatalities as highlighted in each graph in B. (B) Complementary cumulative distribution functions for avalanche scaling variables (distribution of reports in Fig.~\ref{gr:cdf reports}). Points below the lower cutoff in gray. Black lines indicate maximum likelihood fits and error bars 90\% bootstrapped confidence intervals. The data are statistically indistinguishable from power laws at the $p\geq0.1$ significance level except for $P(N)$ (Appendix Fig.~\ref{gr:p-values}) \cite{clausetPowerLawDistributions2009}. (C) Exponents as a function of the separation time $a$. Dashed lines show the average exponent value for the last five points $16\leq a\leq256$\,days.}
\label{gr:fig1}
\end{figure}

We investigate data collected in the Armed Conflict Location \& Event Data Project (ACLED) that aggregates events reported by news media and regional contacts from 1997--2016 \cite{raleighIntroducingACLED2010}. The part of the data set on Africa is notable for its extent---covering two decades, thousands of kilometers, and \mbox{$>10^5$} events. 
We analyze three kinds of events in the data set: Battles involving two or more armed groups (\mbox{$R=42{,}738$}\,reports), Violence Against Civilians in which armed groups attack the population ($R=39{,}127$\,reports), and Riots/Protests ($R=37{,}582$\,reports). Each identified event has geographic coordinate, date, and number of fatalities. Like the canonical avalanche picture for nonequilibrium critical phenomena, we call clusters of events \textit{conflict avalanches}. Although we consider all three conflict types, we focus on the Battles (see Appendices for other event types).

We cluster events into conflict avalanches by setting a separation length $b$ and separation time $a$ such that events that are within the specified distance and time are grouped into the same avalanche (Appendix~B), a procedure analogous to that done for neural avalanches \cite{friedmanUniversalCritical2012,timmeCriticalityMaximizes2016,poncealvarezWholeBrainNeuronal2018}. As we vary these scales, the typical duration and geospatial extent of conflict avalanches change systematically but for a large range of scales observed statistics are remarkably consistent. For the following, we fix $b=140$\,km, allowing large conflicts to percolate while respecting systems boundaries defined by geographic features (e.g., Sahara Desert, coastlines). In Fig.~\ref{gr:fig0}, we show the spatial distribution for the 10 largest avalanches by reports for $b=140\,{\rm km}$ and $a=128\,{\rm days}$. A single example of a conflict avalanche spanning Libya and Tunisia lasting over $10^3$~days with nearly $10^4$ reported fatalities appears in Fig.~\ref{gr:fig1}A along with its temporal profile. Thus, every conflict avalanche has duration $T$ in days, diameter $L$ in kilometers given by the maximally distant pair of events, extent $N$ measured by number of tiling regions or sites, reported fatalities $F$, and number of filed reports $R$. This clustering operation, with only straightforward dependence on physical scales, defines a systematic way of constructing related sets of events in contrast with notions of ``battles'' or ``wars,'' which can depend on sociopolitical nuance.

As visible in Figs.~\ref{gr:fig0} and \ref{gr:fig1}A, the spatial density of conflict is strongly nonuniform. Large conflicts tend to be reported (and perhaps concentrate) along high population areas: few are reported in the Sahara Desert and only a handful are reported in the oceans. Conflict density also appears to depend on factors like geography of country borders (e.g., coastlines, Darfur). These geopolitical features might impose boundaries on conflict propagation. However, communication technology might render physical distance irrelevant for coordinated events. Strong spatial disorder, geographic boundaries, and rapid long distance communication are analogous to effects that destroy scaling in physical systems. Hence, it would be surprising if the length scale $L$ observed in our conflict data fit into a scaling description.

Since such effects are less relevant for time, we use the duration of avalanches $T$ to measure avalanche growth. Then, a minimal scaling hypothesis relating all these properties together predicts
\begin{equation}
\begin{aligned}
	L & \sim T^{1/z}, & N &\sim T^{d_N/z},\\
	F &\sim T^{d_F/z}, & R & \sim T^{d_R/z},\label{eq:dynamical scaling}
\end{aligned}
\end{equation}
with dynamical exponents $1/z$, $d_N/z$, $d_F/z$, and $d_R/z$. The distributions of the scaling variables are likewise expected to scale simply
\begin{equation}
\begin{aligned}
	P(L) &\sim L^{-\nu}, & P(T) &\sim T^{-\alpha}, & P(N) &\sim N^{-u},\\
	P(F) &\sim F^{-\tau},  & P(R) &\sim R^{-\tau'}.\label{eq:scaling}
\end{aligned}
\end{equation}
The relations in Eqs~\ref{eq:dynamical scaling} and \ref{eq:scaling} provide the basis for a scaling hypothesis of armed conflict that we test empirically.

If conflict avalanches grow in time, space, and magnitude in a self-similar manner, we expect the dynamical exponents to be related to the power law exponents. To measure the dynamical exponents, we directly compare the scaling variables to determine $1/z = 0.8\pm0.1$, $d_N/z = 1.4\pm0.1$, $d_F/z = 2.5\pm0.3$, and $d_R/z = 2.0\pm0.3$ (Appendix~C). Then, we construct distributions for each scaling variable independently of one another (Fig.~\ref{gr:fig1}B). 
In all the cases, the power law model is well-aligned with the data, and the data are statistically indistinguishable from power laws in all cases except $P(N)$ \cite{clausetPowerLawDistributions2009}. The corresponding exponents appear in Fig.~\ref{gr:fig1}C. For the highlighted case of $a=128\,$days, we find  $\alpha = 2.44\pm0.13$, $\nu = 2.78\pm0.21$, $u=2.21\pm0.10$, $\tau=1.65\pm0.08$, and $\tau'=1.96\pm0.03$. In a self-consistent framework, measured exponents must satisfy the relations
\begin{equation}
\begin{aligned}
	z(\alpha-1) &= \nu-1 = d_N(u-1)\\
	&= d_F(\tau-1) = d_R(\tau'-1).\label{eq:exp rel 1}
\end{aligned}
\end{equation}
These relations are satisfied within 90\% bootstrapped error intervals. Thus, various features of conflict including, perhaps surprisingly, $L$, are unified in a self-consistent fashion given a simple scaling description.

\begin{figure}[t]\centering
	\includegraphics[width=.95\linewidth]{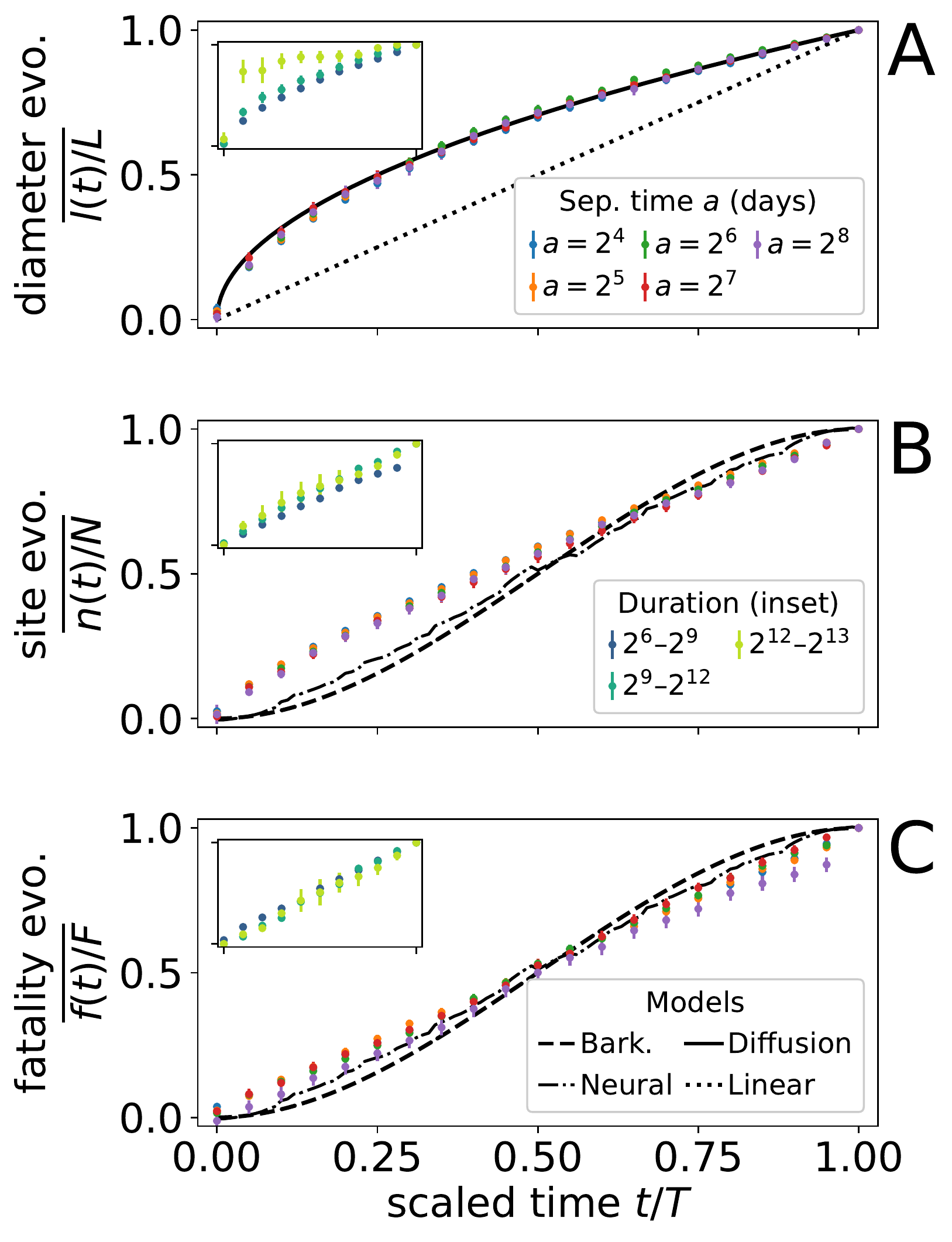}
	\caption{Temporal evolution measured by cumulative fraction of (A) diameter, (B) geographic extent, (C) and fatalities along scaled time $t/T$. Profiles are averaged over all avalanches with duration equal to or exceeding separation time $a$. We compare conflicts with Barkhausen noise  $\int_0^{t/T} V(t')\,dt' = 3(t/T)^2-2(t/T)^3$ (dashed black line \cite{papanikolaouUniversalityPower2011}), experimental neural avalanches (dash-dotted black line, $\rm K>10^3$ \cite{timmeCriticalityMaximizes2016}), and diffusive growth $\overline{l(t)/L} = (t/T)^{1/2}$ (solid black line \cite{randon-furlingConvexHull2009,schutteDiffusionPatterns2011}). 
(insets) For separation time $a=128\,$days, we show averaged profiles after binning by conflict duration and with same axes as main plots. These align except for the few, longest avalanches that quickly saturate in diameter upon reaching geographic or national boundaries. Thus, other effects become relevant when conflict avalanches are commensurate with hard geographic limits. Error bars represent standard errors of the means.}
\label{gr:fig3}
\end{figure}

Self-similarity also predicts that the average evolution of each scaling variable during the course of an avalanche approach a universal profile at large scales. Normalized trajectories of  geographic diameter $\overline{l(t)/L}\sim (t/T)^{1/\zeta}$, extent $\overline{n(t)/N}\sim(t/T)^{\delta_n/\zeta}$, fatalities $\overline{f(t)/F}\sim(t/T)^{\delta_f/\zeta}$, and reports $\overline{r(t)/R}\sim(t/T)^{\delta_r/\zeta}$ give the cumulative fraction of total by scaled time $t/T$ as shown in Fig.~\ref{gr:fig3}. For reports, at least one event occurs at $t=0$ and $t=T$ by construction, so we must account for a $1/R$ ``lattice'' bias to obtain a collapse. A similar bias appears for $n(t)$ and $f(t)$, though stochasticity of the averaged lattice correction for fatalities can induce small negative values at $t=0$ (Appendix~\ref{sec:cum profile}). Having accounted for such biases, we find across avalanches with duration $T\geq a$ that cumulative profiles overlap. This overlap between the temporal profiles indicates that the dynamics of conflict are dominated  by a scale-invariant process as is consistent with a scaling framework.

Notably, the statistical structure encoded in exponent relations in Eq~\ref{eq:exp rel 1} and temporal profiles is largely preserved as we change separation time $a$. In Fig.~\ref{gr:fig1}, we show that the exponents stay close to their values in the highlighted example over an order of magnitude of $16\leq a\leq 256\,$days, and in Fig.~\ref{gr:fig3} the average temporal profiles hardly change across the matching range of $a$. In physical systems, self-similarity is typically explored through symmetries under rescaling. In our case, increasing $a$ groups together events that are increasingly further apart into the same avalanche. Remarkably, we find doubling $a$ is statistically analogous to scaling $T$ in that it largely preserves exponents and temporal profiles across timescales from weeks to years, a result suggesting self-similarity in timing of conflict events \cite{picoliUniversalBursty2015}.

\begin{figure}[h]\centering
	\includegraphics[width=.95\linewidth]{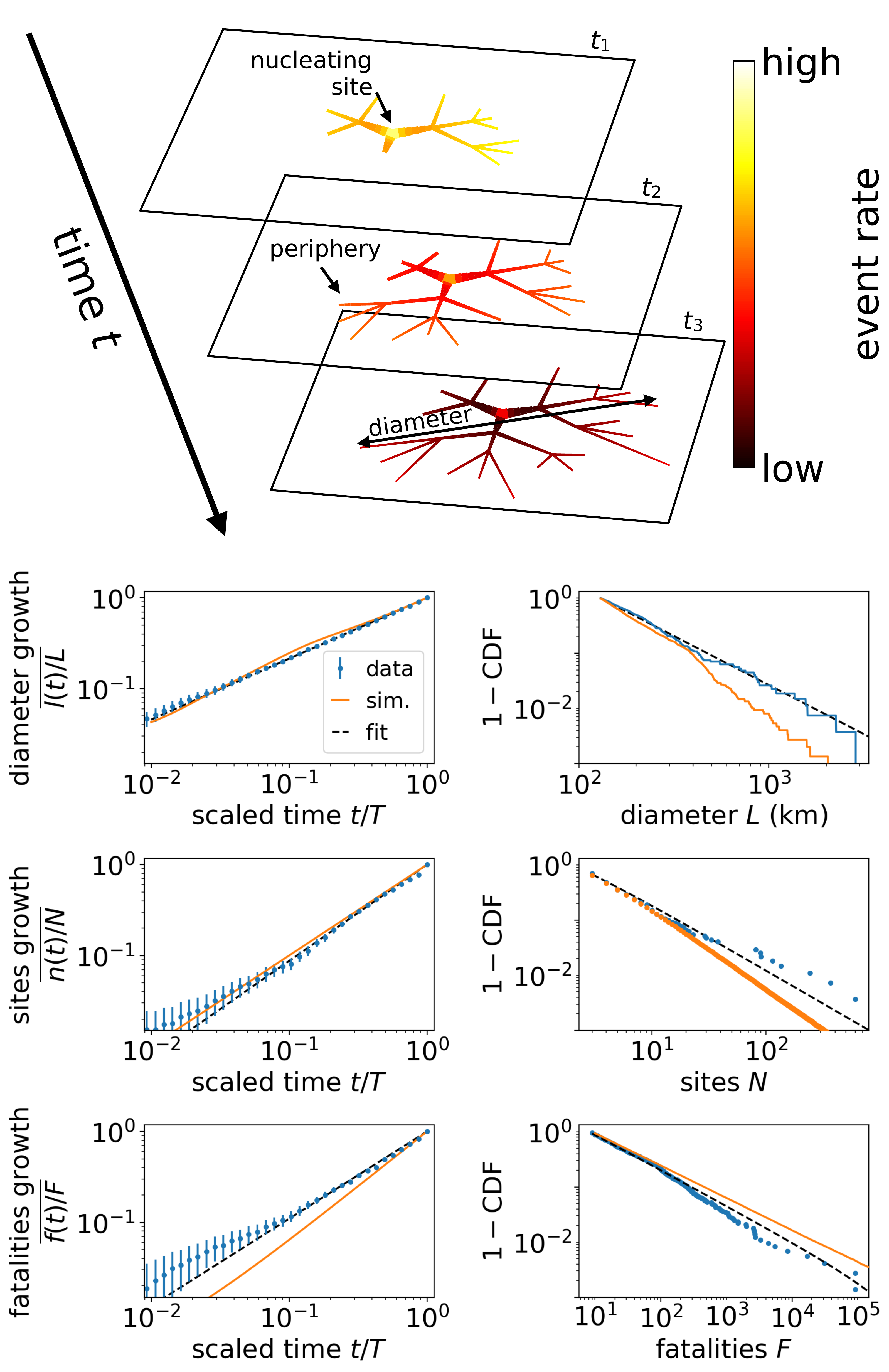}
	\caption{(top) A growing conflict avalanche spreads out to new conflict sites on a fractal, generating conflict events at a slowing rate. As a result, sites near the core tend to have more cumulative events (thick lines) than peripheral sites (thin lines). (bottom) Scaling in data (blue) measured by best fits (dashed black lines) are largely satisfied by the model (orange). Measured dynamical scaling functions are shown after having removed the nonzero intercept at $t=0$ averaged over conflict avalanches with duration $T\geq4$\,days. For $n(t)$, we also require $N>1$ and for $f(t)$ that $F>2$ fatalities. Distributions are shown above lower cutoffs.}\label{gr:conflict model}
\end{figure}

To capture these observations, we propose a randomly branching, armed conflict model, in which conflict spreads from a nucleation site to new, local areas through roads or lines of communication along which goods, people, and ideas move. At each new site, conflict becomes endemic, generating fatalities and reports---though sites at the periphery tend to be less active than in the core. Regional virulence modulates the magnitude of conflict events, affecting the time it takes for conflict to be extinguished. These three components present a minimal set necessary to capture the scaling behavior in armed conflict data, as is further described in reference \cite{leeScalingTheory2020}.

As a first step, we constrain our model using a key property of road networks; that is, intersections become sparse as one moves away from highly interconnected areas \cite{arcauteelsaCitiesRegions2016,kalapalaScaleInvariance2006}. We incorporate this constraint using a randomly branching graph formed by lines of conflict sites of average length $B^k$ that split into an average of $Q$ branches at generation $k+1$ as in Fig.~\ref{gr:conflict model} \cite{burioniFractalsAnomalous1994}. Thus, the locations at which conflict branches split become exponentially further away while conflict sites are distributed with fractal dimension $\delta_n=1+\log{Q}/{\log B} = 1.6$, a value estimated from data. At each point in time, a randomly chosen branch gains one conflict site at unit distance away from its tip such that each branch grows ever slower as conflict spreads. At each conflict site $x_i$, that was reached at time $t_0(x_i)$, fatalities $f_{x_i}(t)$ accumulate such that by time $t\geq t_0(x_i)$,
\begin{align}
	f_{x_i}(t) = v_f(x_i)[t-t_0(x_i)+1]^{1-\gamma_f}[t_0(x_i)+1]^{-\theta_f},\label{eq:fx}
\end{align}
with site virulence $v_f(x_i)$, universal site activity $\gamma_f=0.56$ with 90\% bootstrapped confidence intervals $(0.30,1.00)$, and suppression at distant sites $\theta_f=0.23$ $(0.17, 1.37)$. Total fatalities is a sum over all sites within a conflict avalanche, $f(t) = \sum_i f_{x_i}(t)$. Similar relations hold for reports $r_{x_i}(t)$ and the sum over conflict sites $r(t)$ with $\gamma_r=0.74$ $(0.41,0.90)$ and $\theta_r=0.43$ $(0.35,1.21)$. Thus, these dynamics specify how geographic growth of conflict avalanches in sites $n(t)$ and diameter $l(t)$ is distinct from the social processes for fatalities $f(t)$ and reports $r(t)$.

These dynamics, though consistent with normalized trajectories, do not explain why final conflict magnitude scales superlinearly with dynamical exponents $d_F/z>d_R/z>1 \approx \delta_f/\zeta \approx \delta_r/\zeta$ (Table~\ref{tab:exponents2}). This discrepancy suggests regional variation, captured in average virulence per conflict avalanche $V_r\equiv \br{v_r(x_i)}$, modulates intensity such that some conflicts generate reports at a faster pace. As a result, more virulent conflicts take longer to decay to the report rate threshold set by the separation scales $b$ and $a$, at which point conflict end is determined by the site with the maximum event rate $\partial_t r_{x_i}(t=T)$. This implies a scaling relation we can verify for conflict virulence. In the limit where conflict growth is dominated by rate of reports generated at the core, as implied by $\gamma_r>\theta_r$, a constant rate threshold for conflict extinction implies $V_r \sim T^{\gamma_r}$. If $P(V_r) \sim V_r^{-\beta_r}$, then $\beta_r = (\alpha-1)/\gamma_r+1$, a relation consistent with a fit to the distribution, $\beta_r = 3.0 \pm 0.3$ (details in reference \cite{leeScalingTheory2020}). In summary, universal conflict site growth implies the presence of regional variation that exacerbates conflict intensity beyond what is expected from pure geographic growth. Our model captures such fluctuations driving conflict intensity and provides a quantitative measure in terms of virulence, which may result from multiple forces driving conflict behavior  such as weak governance and level of prosperity \cite{corralFragilityConflict2020}.

Using this set of minimally necessary model components, we simulate conflict avalanches. We closely match observed scaling features as shown in Fig.~\ref{gr:conflict model}.

The emergence of these large-scale symmetries is extraordinary. Such remarkable regularity suggests new directions for prediction of armed conflict dynamics \cite{spagatFundamentalPatterns2018,cedermanPredictingArmed2017}. Our results indicate that temporal profiles can be used to extrapolate from the beginning of a conflict to its end. Scaling relations could aid in estimation of missing data points like fatalities (which are especially difficult to measure) \cite{oloughlinPeeringFog2010,raleighIntroducingACLED2010}, detection of anomalous statistics \cite{durtschiEffectiveUse2004}, or risk assessment for nearby regions by showing how geographic extent scales with duration \cite{salehyanRefugeesSpread2006}.  We capture the confluence of such regularities with a minimal dynamical model inspired by the spread of contagion that aligns closely with the many measured exponents in contrast with models previously compared with armed conflict (Appendix~\ref{sec:other models} and Table~\ref{tab:exponent table}). Unlike canonical cascade models, conflict also includes lattice-site dynamics that evolve with geographic spread. The suppression of these dynamics away from the core could reflect social processes (e.g., ``all politics is local'') or geography (e.g., road density \cite{kalapalaScaleInvariance2006}) that impact conflict evolution. Furthermore, our model suggests conflict is not only the result of local correlations in activity but also of regional and temporal disorder, perhaps reflecting memory of the severity of initiating events \cite{leeCollectiveMemory2017}. We build these conflict avalanches relying only on simple physical scales, propose a quantitative procedure that could complement sociopolitical definition of wars \cite{raleighIntroducingACLED2010}, and capture multiple features of conflict with a low-dimensional description.

Universality in social systems suggests the renormalization group as a powerful means for understanding how physical constraints translate into emergent patterns at large scales \cite{fortunatoScalingUniversality2007,bettencourtGrowthInnovation2007}. While social and biological systems are complicated by the role of information-processing and are limited in experimental control \cite{flackLifeInformation2017,flackCoarsegrainingDownward2017}, ordering and pattern formation are ubiquitous \cite{krakauerBetterLiving2010,krakauerChallengesScope2011}. Our findings hint at physical principles that unify their seemingly diverse and contingent properties.

\begin{acknowledgments}
We thank Veit Elser, Guru Khalsa, Jaron Kent-Dobias, and Simon Dedeo for helpful discussion. We acknowledge an NSF GRFP under grant no. DGE-1650441 (EDL), NSF no. 0904863 (JCF \& DCK), a St. Andrews Foundation grant of no. 13337 (EDL, JCF, \& DCK), a John Templeton Foundation grant of no. 60501 (JCF \& DCK), the Proteus Foundation (JCF), and the Bengier Foundation (JCF).
\end{acknowledgments}

\bibliographystyle{unsrt}
\bibliography{refs}

%
%
%
%
%
%
%

\clearpage
\appendix
\setcounter{secnumdepth}{2}
\renewcommand{\thefigure}{S\arabic{figure}}
\setcounter{figure}{0}
\renewcommand{\thetable}{S\arabic{table}}
\setcounter{table}{0}
\renewcommand{\thesection}{\Alph{section}}
\setcounter{section}{0}

\section{Armed Conflict Location and Event Data (ACLED) Project}\label{sec:data}
We use the data set provided online as ACLED Version 7 \citeapp{raleighIntroducingACLED2010}. This project measures political violence around the world with a focus on African states for 20 years (Jan. 1, 1997 through Dec. 31, 2016). The data set is organized around events, which have a specific date and time. We analyze three types of events included in the data set: Battles between armed groups (\mbox{$K=42{,}738$}), Violence Against Civilians ($K=39{,}127$), and Riots/Protests ($K=37{,}582$). 

\begin{figure}[b]
	\includegraphics[width=.9\linewidth]{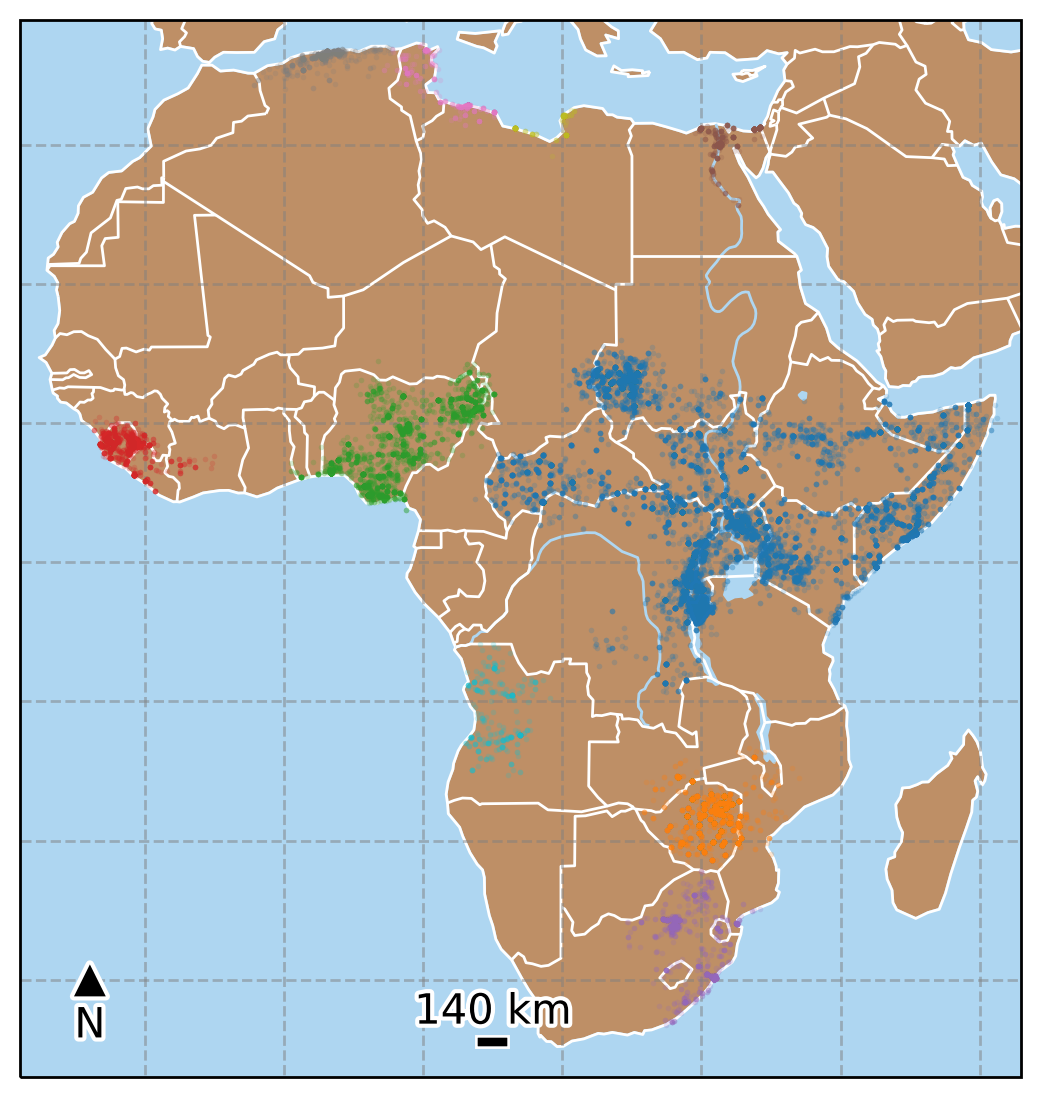}
	\caption{Spatial distribution of 10 largest conflicts involving Violence Against Civilians (VAC) given $b=140\,$km and $a=128\,$days. Map made with Natural Earth.}
\end{figure}

According to the codebook, there are three different kinds of battles that we include in our Battles conflict avalanches. As quoted from the codebook, these are defined as follows:
\begin{enumerate}
	\item Battles - No change of territory: ``A battle between two violent armed groups where control of the contested location does not change. This is the correct event type if the government controls an area, fights with rebels and wins; if rebels control a location and maintain control after fighting with government forces; or if two militia groups are fighting. Battles take place between a range of actors.''
	\item Battle - Non-state actor overtakes territory: ``A battle between two violent armed groups where non-state actors win control of a location. If, after fighting with another force, a non-state group acquires control, or if two non-state groups fight and the group that did not begin with control acquires it, this is the correct event. There are few cases where opposition groups other than rebels acquire territory.''
	\item Battle - Government regains territory: ``A battle between two violent armed groups where the government (or its affiliates) regains control of a location. This event type is used solely for government re-acquisition of control. A small number of events of this type include militias operating on behalf of the government to regain territory outside of areas of a government's direct control (for example, proxy militias in Somalia which hold territory independently but are allied with the Federal Government).''
\end{enumerate}

\begin{figure}[b]
	\includegraphics[width=.9\linewidth]{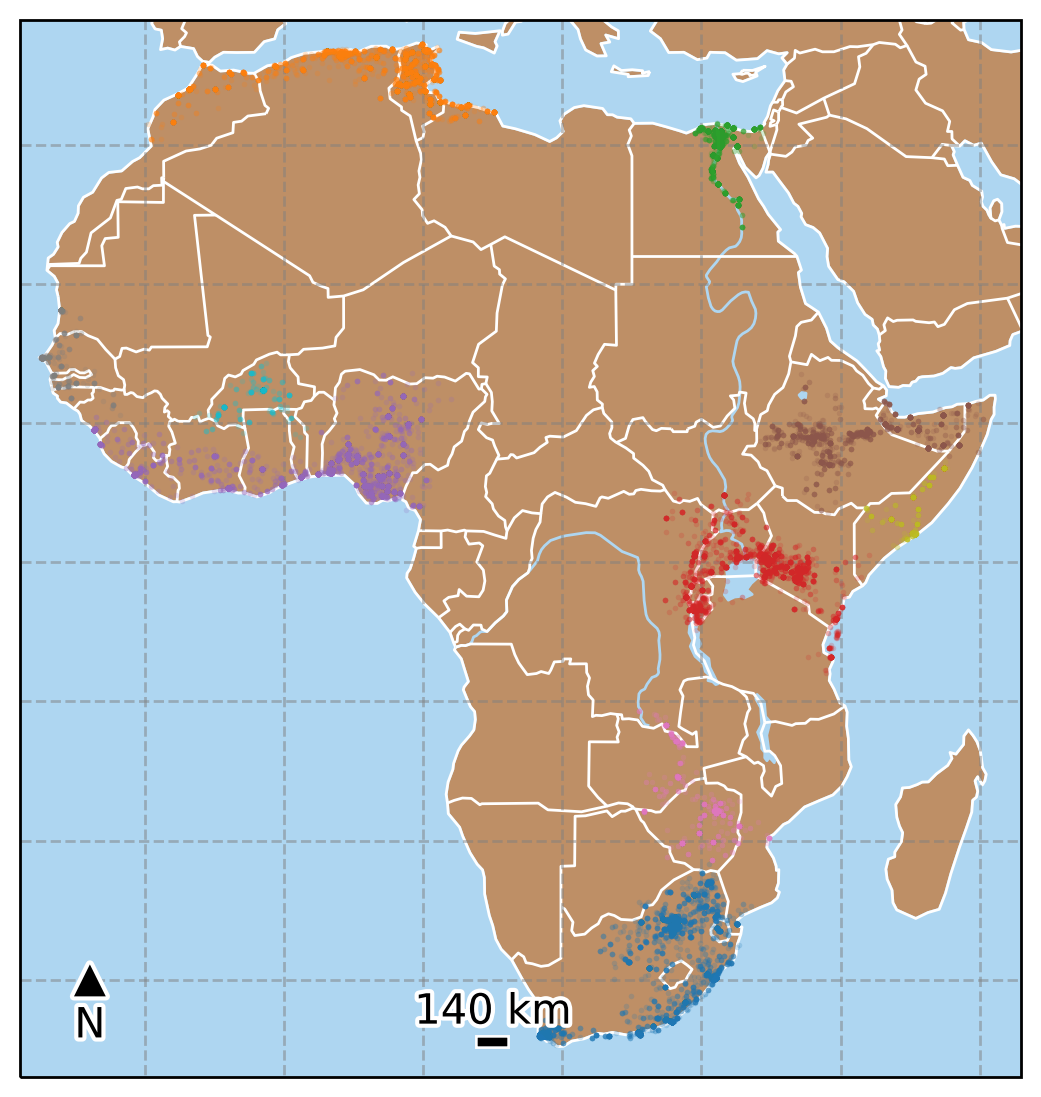}
	\caption{Spatial distribution of 10 largest conflicts involving Riots/Protests given $b=140\,$km and $a=128\,$days.}
\end{figure}

We also investigate Violence Against Civilians (VAC):
\begin{quotation}
Violence against civilians is a violent act upon civilians by an armed, organized, and violent group. By definition, civilians are unarmed and not engaged in political violence. Rebels, governments, militias, external forces, and rioters can all commit violence against civilians. Protesters are also civilians, and significant violence against protesters falls under this category.
\end{quotation}

\begin{figure}[tb]
	\includegraphics[width=\linewidth]{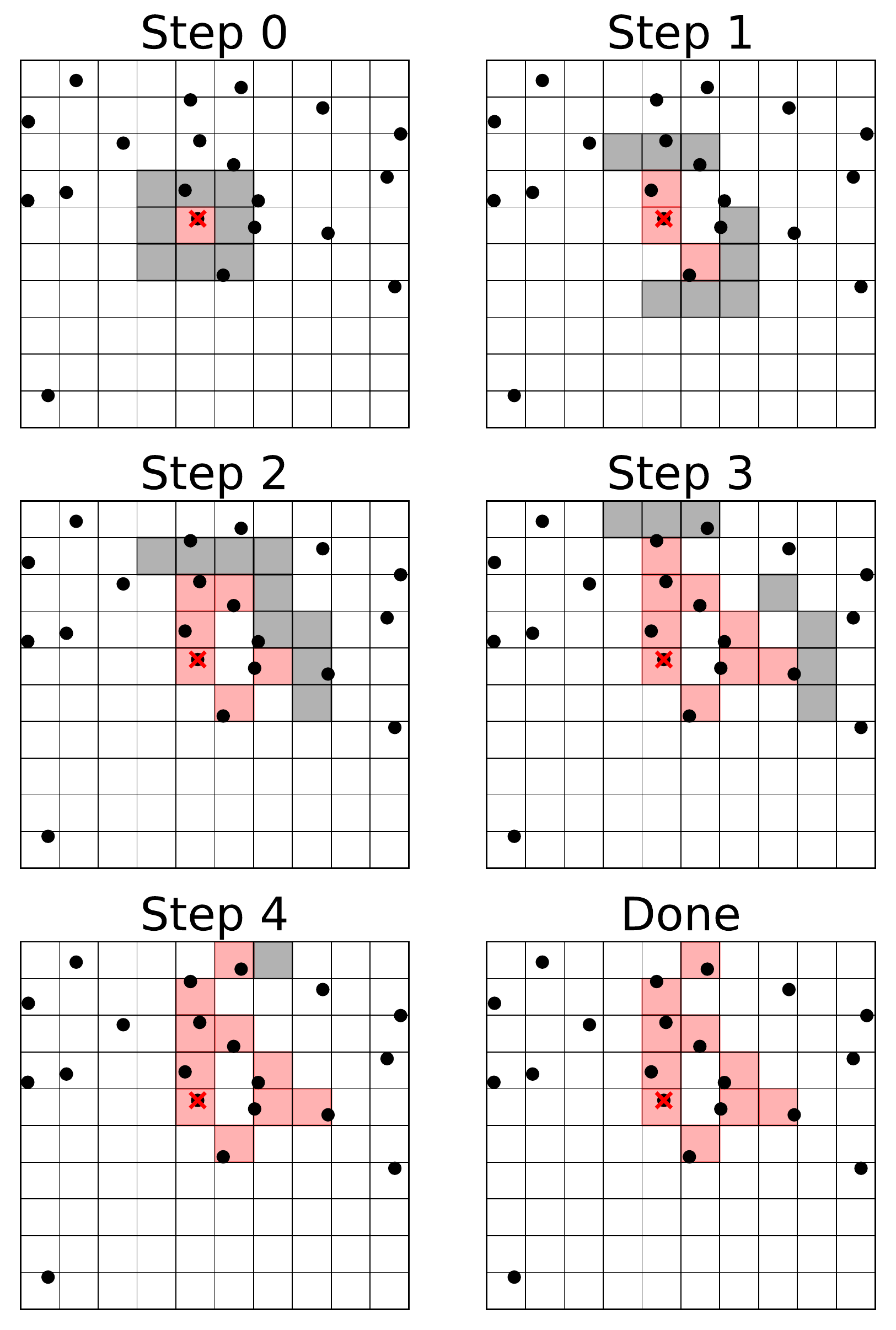}
	\caption{Schematic of clustering algorithm for building a conflict avalanche. At Step 0, the algorithm picks a random event and begins building a cluster there. Then, all new neighbors (gray) of new tiles added to the cluster (red) are evaluated for events (black circles). When no more tiles can be added, the algorithm stops.}\label{gr:cluster algo}
\end{figure}

Finally, there are Riots/Protests:
\begin{quotation}
A protest is a public demonstration in which the participants do not engage in violence, though violence may be used against them. Often---though not always---protests are against a government institution. Rioting is a violent form of demonstration where the participants engage in violent acts, including but not limited to rock throwing, property destruction, etc. Both of these can be coded as one-sided events. All rioters and protesters are noted by generic terms (e.g. `Rioters (Country)' or `Protesters (Country)'); if representing a group, the name of that group is recorded in the respective `associated actor' column.
\end{quotation}

In the analysis, we only consider statistics of the conflict avalanches where $T>1$, $R>1$, $F>1$, and $L>0$ although an event does not have to satisfy all cutoffs simultaneously, i.e., we may use it for $P(R)$ but not $P(F)$. We find that the statistics of events below these lower cutoffs generally deviate from the observed power law statistics in the rest of the distribution, and such deviations are likely attributable to data problems. For some events, ACLED sets the estimate of fatalities $F=0$ unless they have confirmed with a ``reputable source,'' so some of these cases are simply missing statistics (there is no way to distinguish between missing data or no fatalities).\footnote{Accurate data on conflict is difficult and even dangerous to collect and necessarily this data set does not sample all events with equal accuracy or detail. Nevertheless, a conflict data project of this scale is unprecedented.}
As for time scales, the highest precision available in the data set is to the day which defines a lattice scale below which we cannot probe. As for length scales, we find many events occur exactly at the same geographic coordinates which presumably also involve some lattice scale below which the data aggregators either could not access or did not find a pressing need to do so. Such resolution effects are akin to rounding artifacts common in human reported data like the Iraq War Logs where reported times are rounded to 10 or 30 minute intervals. Importantly, these anomalies matter little at large scales where such effects are dominated by the aggregate regularities of the system.

\begin{figure*}[tb]
	\includegraphics[width=.9\linewidth]{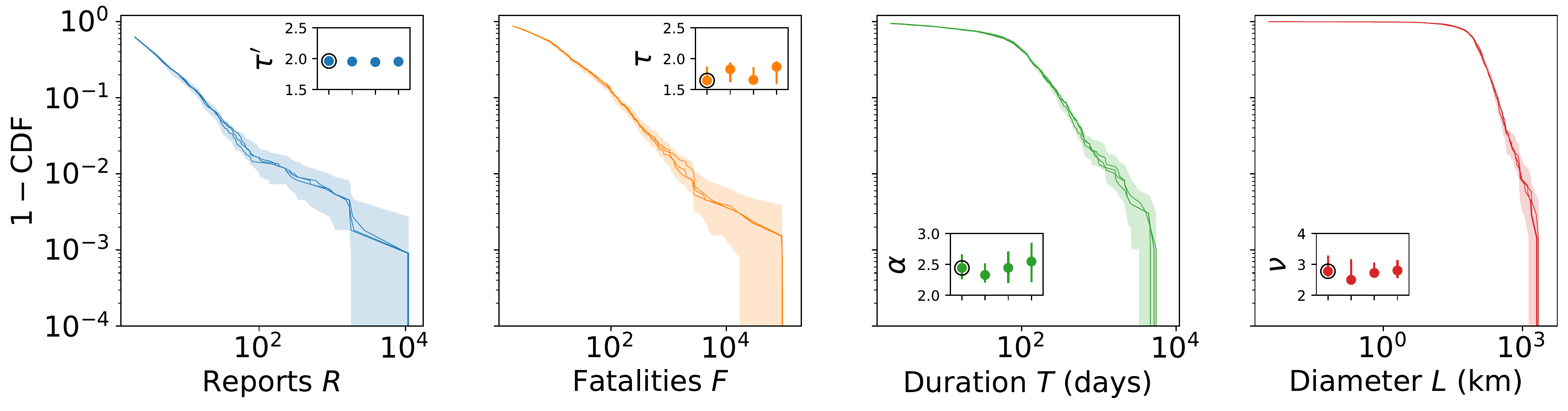}
	\caption{Distributions of scaling variables for several random Voronoi tilings given by the complementary cumulative distribution function (CDF). Bootstrapped confidence intervals of 90\% for the data in Fig.~2B are shown behind the distributions for three other Voronoi tilings (discrete points connected by lines for visibility). These all fall well within expected statistical variation. (inset) Measured exponents for the four distributions where the circled exponent corresponds to data used in the main text. The choice of tiling does not substantially alter the exponent. The fluctuations in the means visible, for example for $\tau$, reflect variation in the lower bound found for the data, variability that is inherent in the fitting procedure when a single lower bound is chosen \protect\citeapp{clausetPowerLawDistributions2009}. Importantly, this fluctuation is captured by bootstrap confidence intervals.}
	\label{gr:voronoi cdfs}
\end{figure*}

\section{Clustering algorithm}\label{sec:clustering}
To generate our conflict avalanches, we choose a separation length scale $b$ and separation time scale $a$ that correspond to the minimum separation between sequential pairs of events in a single avalanche. To do this, we first bin the time points into bins of width $a$ and consider any contiguous sequence of bins with at least one event to be potentially (we must account for geographic distance next) part of the same conflict avalanche. In contrast to how avalanches are constructed for neural systems \citeapp{timmeCriticalityMaximizes2016}, we do not discretize the day on which avalanches occurred to the scale $a$ after constructing the avalanche, but preserve the precise time at which events were reported. Such discretization to a lattice scale is unnecessary for exploring scaling relations. As a result, the temporal clustering procedure constructs sequences of contiguous events where breaks are inserted between any pair of events with at least separation of $a$ days.

An exact analog of this unidimensional procedure to the surface of the Earth is impossible because no regular tiling of the surface of a sphere exists. Surely, one approach without bins would be to measure directly the pairwise distance between every pair of events, but this approach scales quadratically and is particularly slow because geodesic distance calculations are expensive. With our data set of $10^4$--$10^5$ events, such a procedure would take inordinately long on a desktop computer. Instead, we generate a Voronoi tiling of the Earth using a Poisson disc-sampling algorithm to generate a random but regularly-spaced set of tiles with average spacing of $b/2$ \citeapp{daviesPoissonDiscSampling}. Neighboring ``bins'' correspond to Voronoi tiles whose centers are within a fixed distance $b$, and we can search for contiguous sets of tiles that have at least one event.\footnote{More generally, a tiling with spacing $b/k$ has resolution (and computational cost) that increases with $k$. Larger $k$ reduces variability amongst different random Voronoi grids and when $k\rightarrow\infty$, there is a unique clustering equivalent to calculating the pairwise distance between every pair of events. For our data, we find that $k=2$ is sufficient to return similar statistics between different Voronoi grids for $b=140$\,km.} Importantly, this Voronoi algorithm only involves distance calculations that scale as the square of the number of {\it tiles} regardless of the density of events.

\begin{figure}
	\includegraphics[width=.49\linewidth]{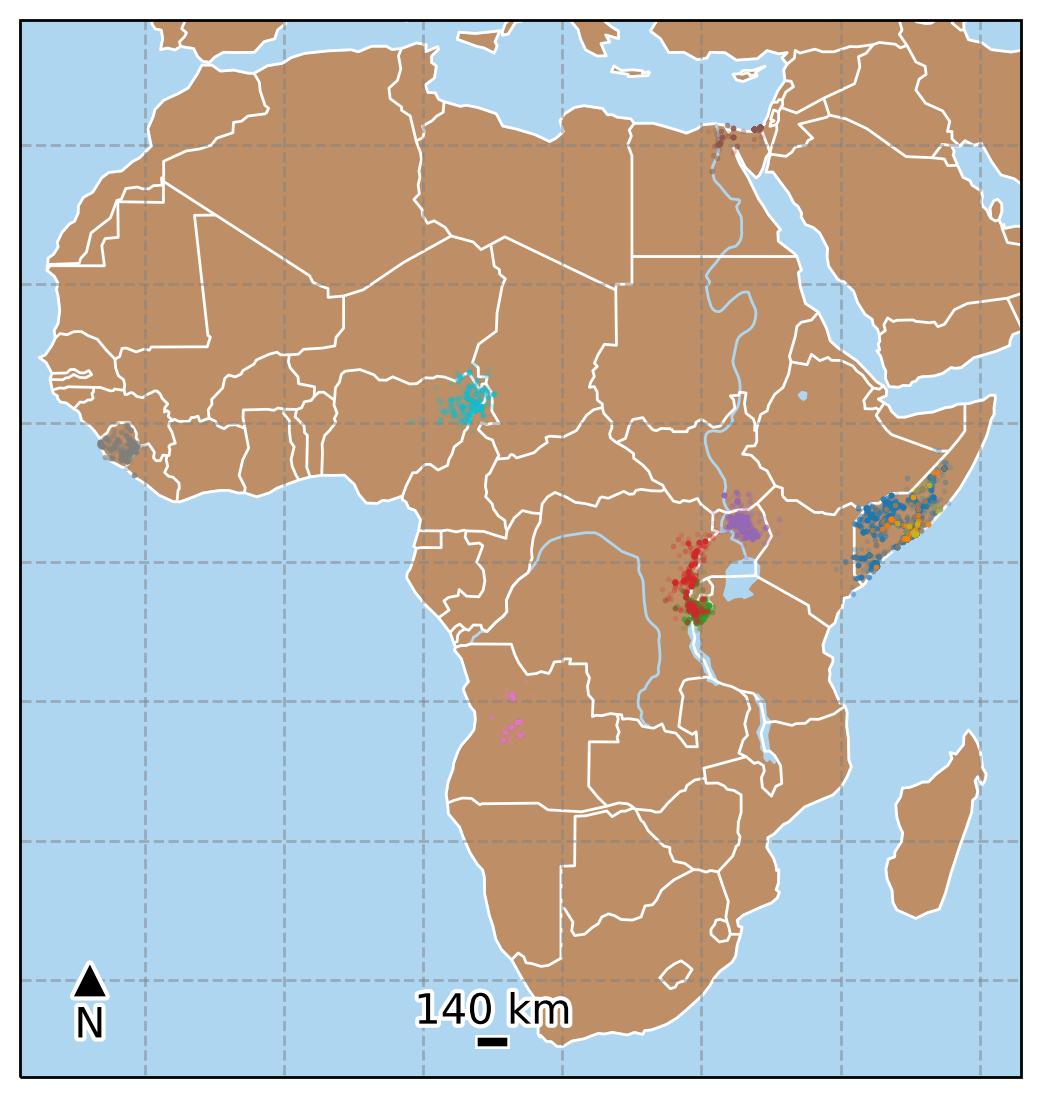}\includegraphics[width=.49\linewidth]{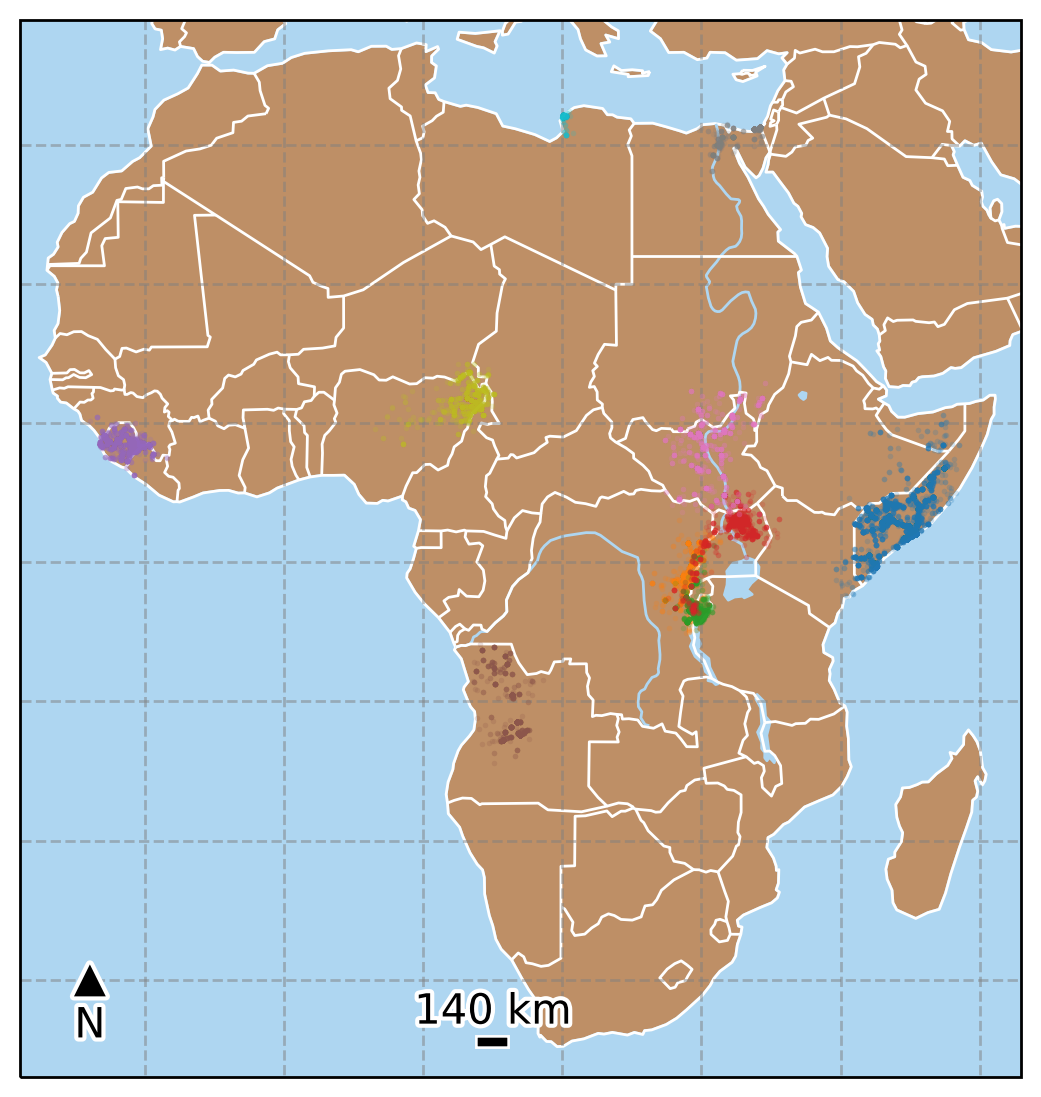}
	\includegraphics[width=.49\linewidth]{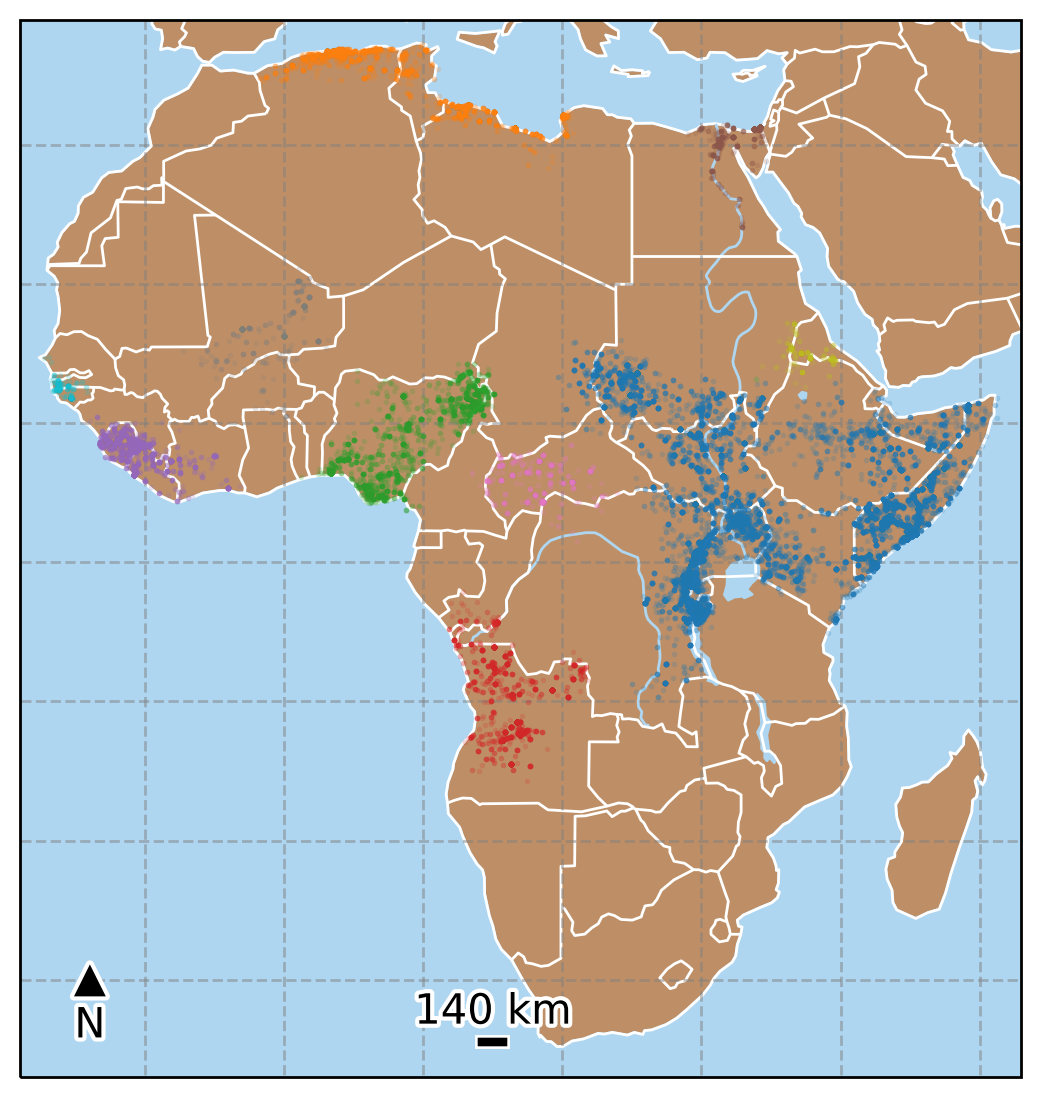}\includegraphics[width=.49\linewidth]{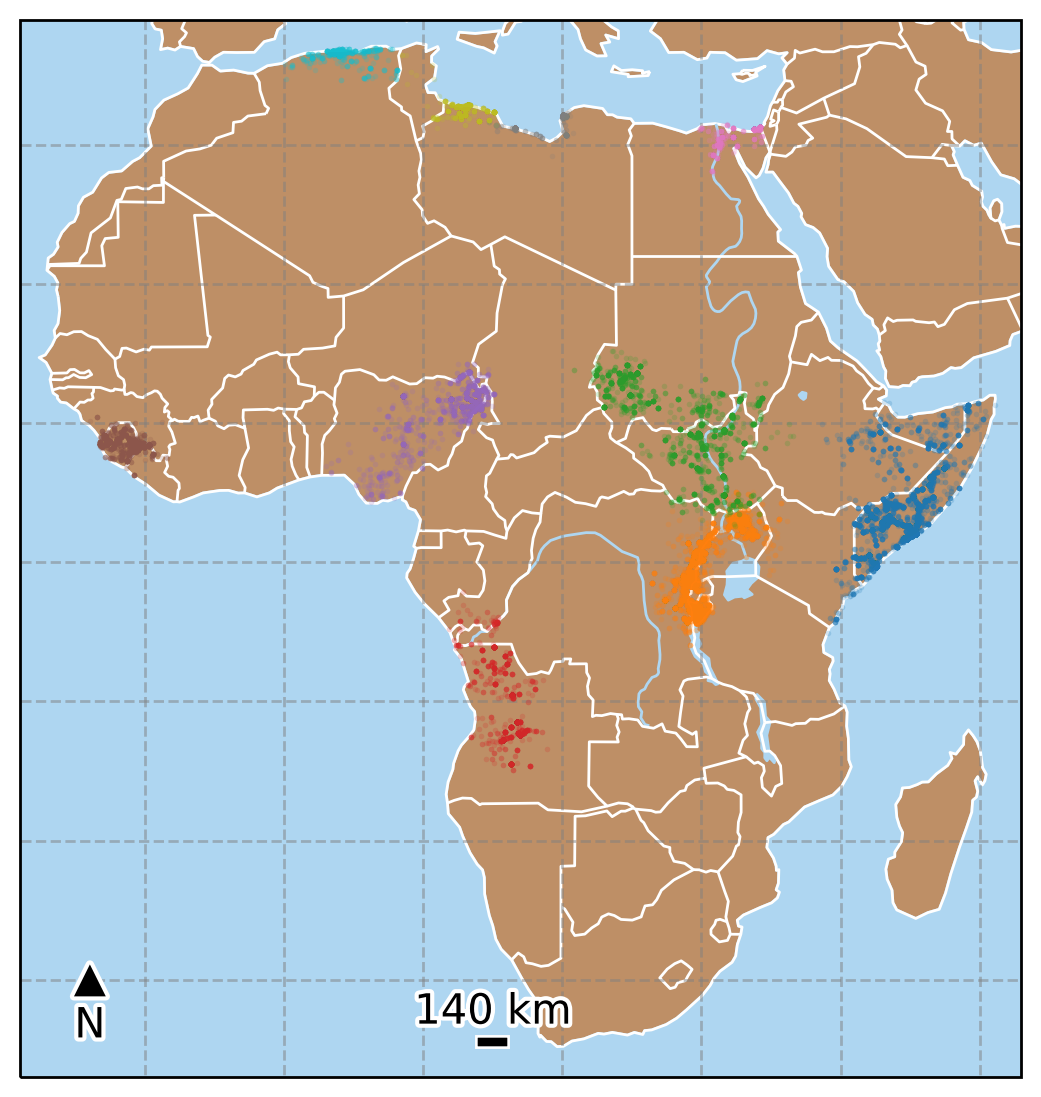}
	\caption{Spatial distribution of 10 largest Battle conflict avalanches for $b=140\,$km and multiple separation scales $a$. (clockwise from top left) $a=16$\,days, $a=32$\,days, $a=64$\,days, $a=256$\,days.}
	\label{gr:map of multiple time scales}
\end{figure}

\begin{figure*}[tbp]
	\includegraphics[width=\linewidth, clip=true, trim=0 0 0 0]{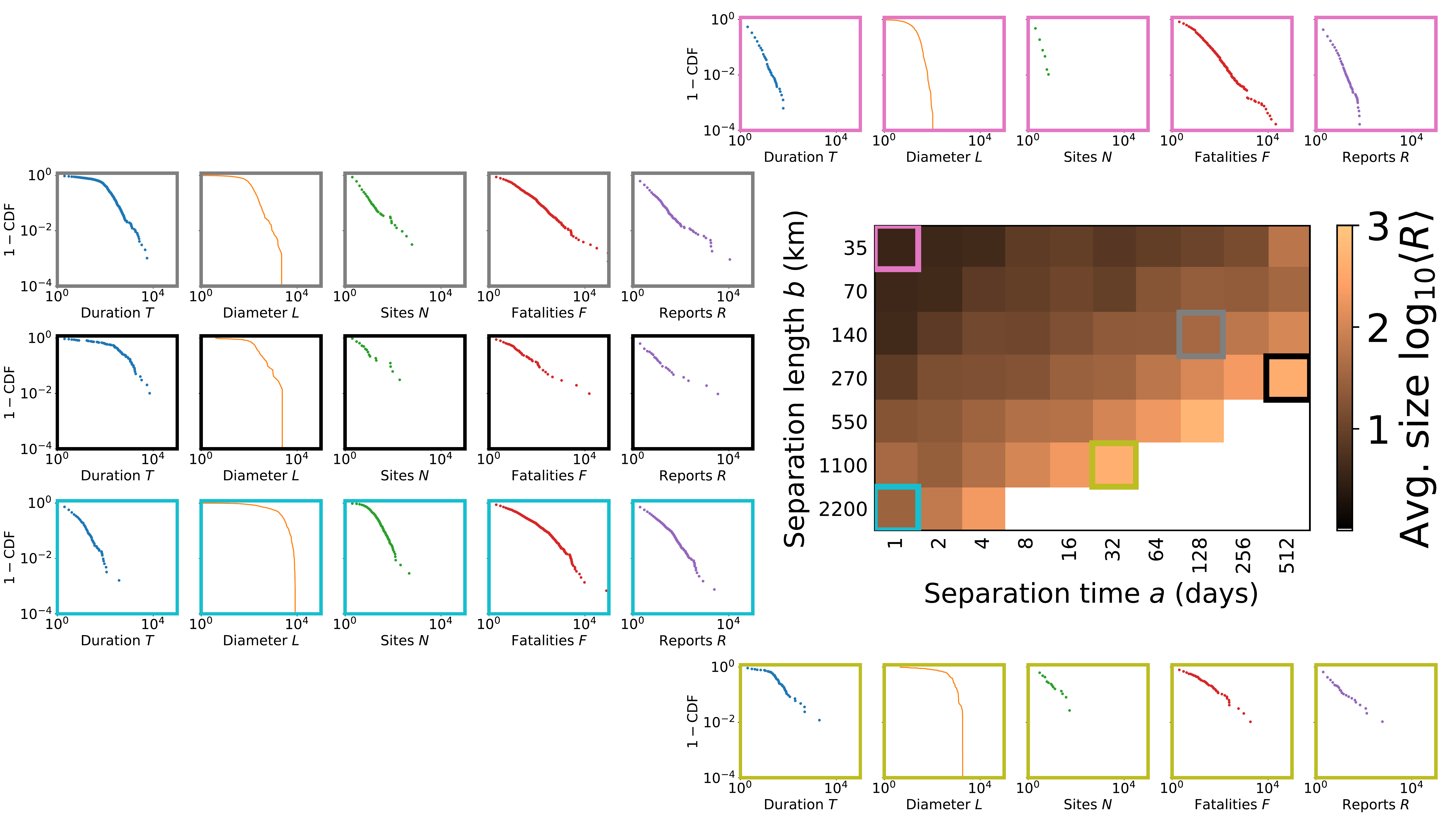}
	\caption{Distributions of scaling variables $T$, $L$, $N$, $F$, and $R$ across a range of spatiotemporal scales $35\,{\rm km}\leq b \leq 2,200$\,km and $1\,{\rm day}\leq a\leq512$\,days for Battles. (center) We show the average avalanche size as reports $\br{R}$ given a clustering spatiotemporal scale ($b,a$) to give a sense of the variation across all scales. Where we have $\rm K<50$ data points above the lower cutoff, the region is whited out.  (top, pink) When $b$ is small, avalanches are likewise small ($R<10^2$ including the largest observed avalanche) and show little dynamic range ($T<10^2$). (top left, gray) In a middle range of $b$, conflict avalanches exist for a wide range of scales, corresponding to the data that we analyze in the main text. (middle left, black) When the time scale $a$ is comparable to the total duration of the data set ($\sim 8{,}000$\,days), avalanches approach the spatial and temporal limits of the data, we have many fewer avalanches to examine, and so we lose dynamic range in the statistics. (bottom left \& right, teal \& gold) When the separation length $b$ is comparable to the entire extent of the African continent ($\sim8{,}000$\,km), most conflict avalanches span the system as visible from the diameters $L$, and conflict avalanches are few.}\label{gr:scale range}
\end{figure*}

\begin{figure}[tb]
	\includegraphics[width=\linewidth]{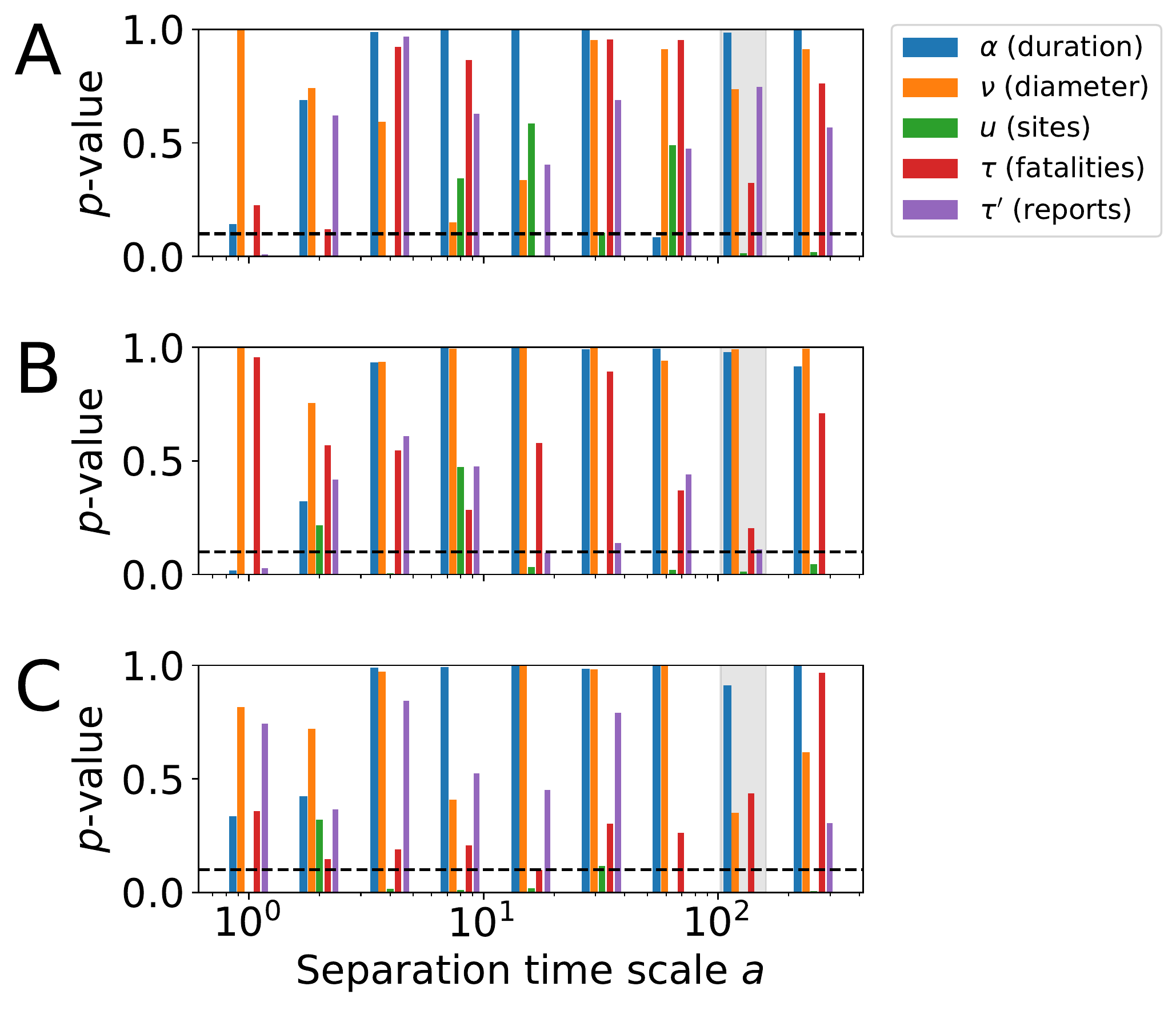}
	\caption{Results of statistical tests for power law fits to (A) Battles, (B) VAC, and (C) Riots/Protests. We consider distributions with $p\geq0.1$ to be statistically indistinguishable from power laws \protect\citeapp{clausetPowerLawDistributions2009}. The exponents for the separation time we use in the main text $a=128$\,days are highlighted in gray.}
	\label{gr:p-values}
\end{figure}

As a simple demonstration of the algorithm, we provide a schematic in Fig.~\ref{gr:cluster algo} that iterates through the construction of a single conflict avalanche in a 2-dimensional space (or one dimension of space and one of time). In this particular example, each tile has exactly 8 neighbors, whereas the actual number of neighbors will vary randomly in the Voronoi tiling. At each step, all {\it new} closest neighbors (gray) of the cluster (red) are evaluated and appended onto the existing cluster if they contain an event (black point). Once the cluster can no longer grow because there are no neighboring tiles with events, the algorithm stops. This procedure defines a systematic way of constructing sequences of related events given spatial and temporal scales.

Although different random Voronoi tilings will cluster events in a slightly different way, we find that the variation from such randomness is small compared to the statistical variation estimated from bootstrapped confidence intervals for a single Voronoi tiling. As we show in Fig.~\ref{gr:voronoi cdfs}, the distributions of conflict statistics across several random Voronoi tilings are all very similar. The measured exponents likewise agree within the bootstrapped confidence intervals. Thus, the Voronoi clustering procedure serves as a computationally efficient way of generalizing the temporal discretization procedure used to identify contiguous events in one dimension to curved surfaces. 

\begin{figure}[tb]
	\includegraphics[width=\linewidth]{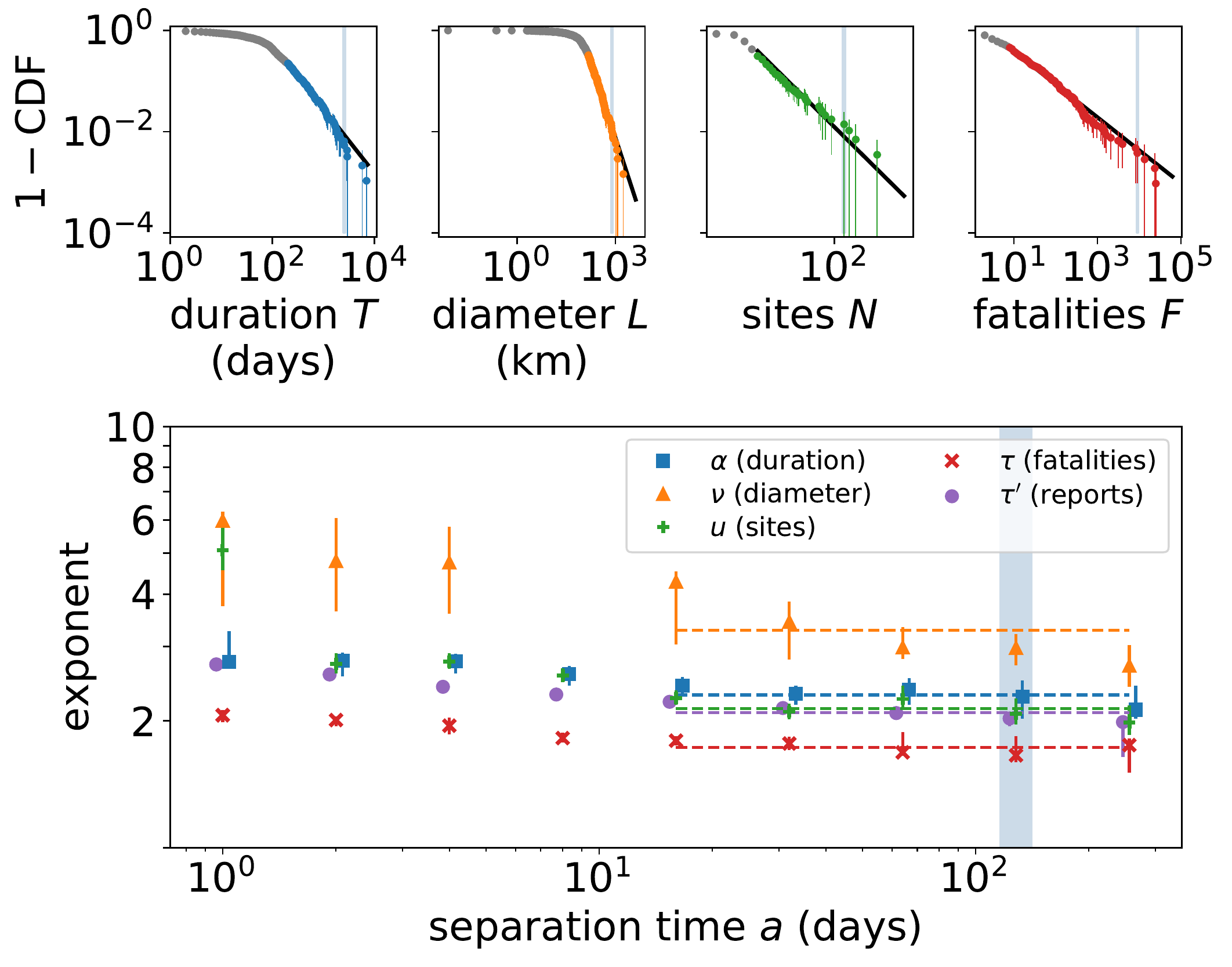}
	\caption{Measured exponents for VAC. All shown distributions for $a=128$\,days are indistinguishable from power law distributions at the $p\geq0.1$ level according to the KS test. There is a missing point for $\nu$ at $a=16$\,days because the measured value exceeds the upper 90\% confidence bound. Such an artifact can occur when the tail of the distribution is not sampled well as can happen with a large lower cutoff. In these cases, the measured exponent may be unreliable.}\label{gr:vac exponents}
\end{figure}

\begin{figure}[tb]
	\includegraphics[width=\linewidth]{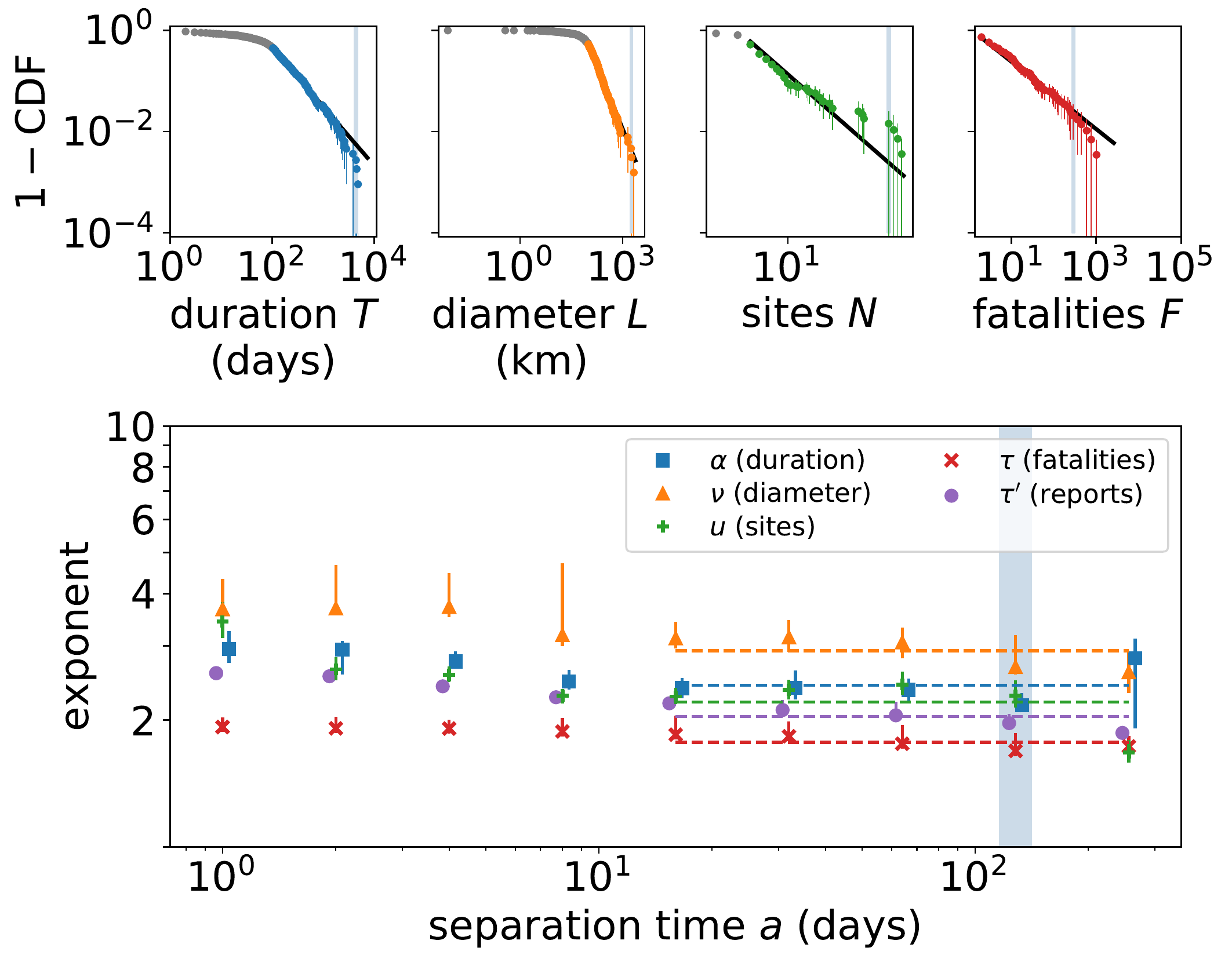}
	\caption{Measured exponents for Riots/Protests. All shown distributions for $a=128\,$days are indistinguishable from power laws at the $p\geq0.1$ significance level according to the KS test except for $P(R)$. Yet, the separation times nearby $a=128\,$days, namely $a=32$\,days and $a=256\,$days, serve as bounds on the possible bias of the exponent estimate. Given that the bound is tight and continuing with the best estimate of the exponent, the scaling relations specified in Eq~\ref{eq:scaling} are satisfied.}\label{gr:riots exponents}
\end{figure}

\begin{figure*}[tb]
	\includegraphics[width=\linewidth]{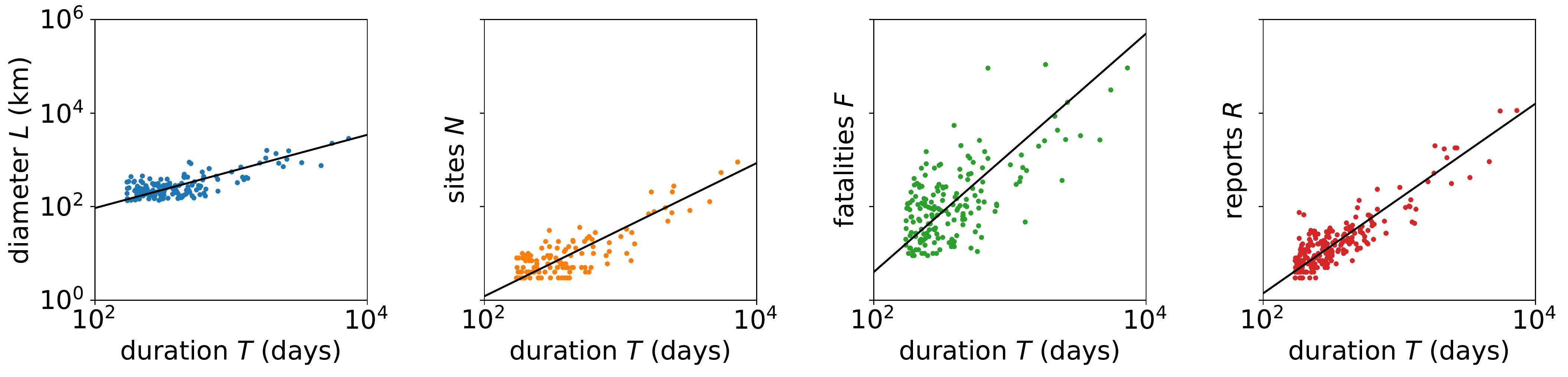}
	\caption{Measurement of Battles dynamical exponents $1/z$, $d_N/z$, $d_F/z$, and $d_R/z$ by minimization of Eq~\ref{eq:log cost}. Only values above the lower bounds (found from fitting the distributions in Fig.~2) are fit. We find $1/z = 0.8\pm 0.1$, $d_N/z = 1.4\pm0.1$, $d_F/z = 2.5 \pm 0.3$, and $d_R/z = 2.0 \pm 0.3$.}\label{gr:fractal reg}
\end{figure*}

As we mention in the main text, we focus on \mbox{$b=140\,$km} because it presents a ``Goldilocks'' zone where avalanches occur over a wide range of sizes. In Fig.~\ref{gr:map of multiple time scales}, we show the spatial distribution of the largest 10 Battles conflict avalanches as we vary $a$ with $b=140\,$km fixed as in the main text. In Fig.~\ref{gr:scale range}, we present an overview of avalanche statistics across a much broader range of spatiotemporal scales $(b,a)$ ranging from $35\,{\rm km}\leq b \leq 2200$\,km and $1\,{\rm day}\leq a\leq512$\,days. When the temporal scales are short, avalanches do not percolate far and we are limited to very small, short, and spatially localized conflicts (pink box representing $b=35\,$km and $a=1\,$day). Although most of the variables here show limited dynamic range, the distribution of fatalities is spread out across three orders of magnitude. The fact that fatalities are heavy-tailed both with and without accounting for spatiotemporal scales may explain why fatalities are so prominent in the armed conflict literature. When we go to much larger scales of $b=1{,}080$--$2{,}060$\,km and $a=512$\,days (black, teal and gold boxes), a few avalanches start to span the physical size of the African continent ($\sim 8{,}000$\,km) and the time series ($\sim8{,}000$\,days). We would expect boundary effects to dominate in this regime and correspondingly avalanche space and time scales are compressed to a small region along the system cutoffs. As a result, we have many fewer conflict avalanches on which to estimate scaling parameters, so we avoid this regime. For a middle range of $b$ around $10^2$\,km, we can probe a wide range of temporal scales for avalanches that display scale-invariant statistics in the tails while also accumulating a reasonable number of temporal profiles to evaluate. 

Indeed, the choice of appropriate scale on which to define related events is a problem that has received much attention in the context of neural avalanches. For neural avalanches, researchers must determine appropriate interspike intervals and often must account for a fixed electrode spacing while recording from a sparse sample of a neural culture \citeapp{beggsNeuronalAvalanches2003}---although new high-resolution, nearly single-cell optical techniques have become possible \citeapp{poncealvarezWholeBrainNeuronal2018}. In principle, the physical layout of axonal and dendritic connections determines a causal network for neural spike propagation and so direct measurement of true (not only statistical) sequences should be possible. In practice, such measurements are not yet feasible and spatiotemporal proximity is often used as a proxy where a good rule-of-thumb is the average interspike interval as a measure of characteristic time scale. When electrode arrays that effectively define a coarse grid are used, the time scale defining related events must be scaled with the distance between the electrodes because the finite propagation velocity of neural signals sets a relevant scale \citeapp{beggsNeuronalAvalanches2003}. Furthermore, other statistical techniques for detecting causality have been explored for constructing ``causal networks'' that induce very different distributions \citeapp{williams-garciaUnveilingCausal2017}. Such techniques for determining networks of related events present an opportunity for further work in armed conflict avalanches.

For our work, sociopolitical information could be used to cluster events into familiar notions of battles or wars, but such clustering is not deterministic and includes ambiguity both in identification of actors and attribution of responsibility \citeapp{raleighIntroducingACLED2010}. 
Nevertheless, our algorithm for building conflict avalanches generates ones that align with sociopolitical inuition. One example is the one that we provide in Fig. 2. It is the spread of unrest that started in Tunisia and subsequently flared up in neighboring Libya as part of the Arab Spring \citeapp{lynchArabUprising2013}. These events were recognized as two separate revolutions occurring in two different countries, but given their overlap in space and time might be labeled ``Beginning of the Arab Spring.'' Another particularly large avalanche we find involves events occurring in both Angola and Congo. According to historical record, these events occurred during the Angolan Civil War (1975--2002) and the Second Congo War (1983--2005), which despite separate labels are known to be closely intertwined \citeapp{AngolanCivil,SecondCongo,raleighIntroducingACLED2010}. Another large set of events that contain two large confrontations is a combination of the Second Sudanese Civil War (1983--2005) and the Eritrean-Ethiopean War (1998--2000) \citeapp{EritreanEthiopian,SecondSudanese,raleighIntroducingACLED2010}. This latter example is one where two sociopolitically distinct events happened in close enough proximity to be joined into a single large avalanche. Such combinations become increasingly frequent as the separation length $a$ increases, but as we show in the main text such concatenation generates longer conflict avalanches that remained aligned with our scaling framework. Finally, we point out that the origins and relations between conflict events is actively disputed, so our definition provides an alternative description that serves as a systematic measure along which to compare otherwise defined conflict aggregates. We take the simplest (and neutral) approach where correlations in cause can be imputed to {\it physical} spatiotemporal proximity, leading to the surprising conclusion that the spread of armed conflict might be described in the language of critical phenomena.

\begin{table*}
	\includegraphics[width=.62\linewidth,clip=true]{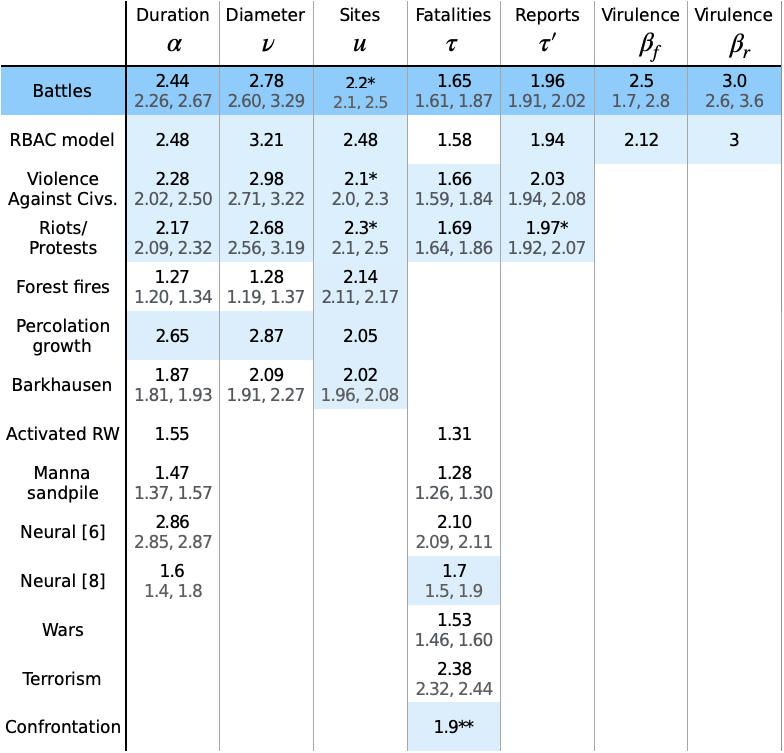}
	\caption{Table of distribution exponents for the data, our RBAC model, and other models and data sets. See Table~\ref{tab:exponents2} for dynamical exponents. Light blue indicates overlap with confidence intervals measured for Battles. Cases where the power law model is significantly different from the data given our $p\geq0.1$ threshold, we indicate with an asterisk. Here, we include two examples of neural avalanches from zebrafish \protect\citeapp{poncealvarezWholeBrainNeuronal2018} and cortical culture \protect\citeapp{friedmanUniversalCritical2012}, terrorism \protect\citeapp{clausetFrequencySevere2007}, and a coalescence-fragmentation model applied to confrontation \protect\citeapp{johnsonSimpleMathematical2013}. For the latter model, this exponent can be found for confrontation on a two-dimensional grid, though the power law fit is significantly different from the model distribution. For conflict avalanches, the uncertainty range corresponds to 90\% bootstrapped confidence intervals. For the other examples, we take the error bars directly from the cited work. None of these other models have exponents that align closely with armed conflict dynamics across multiple features with the surprising exception of percolation growth, which is described further in Appendix~\ref{sec:other models}.}
	\label{tab:exponent table}
\end{table*}

\begin{table*}
	\includegraphics[width=.9\linewidth,clip=true]{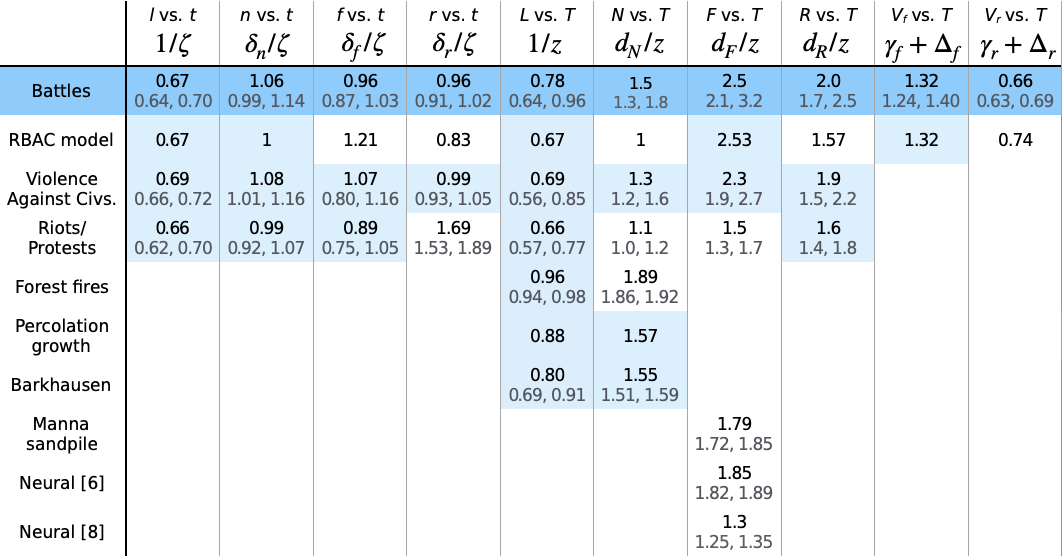}
	\caption{Table of dynamical exponents. See Table~\ref{tab:exponent table} for other exponents and description.}\label{tab:exponents2}
\end{table*}

\section{Power law fitting}
Given the conflict avalanches for a given length scale $b$ and time scale $a$, we extract the scaling variables $R$, $F$, $T$, and $L$ to measure the distribution exponents $\tau'$, $\tau$, $\alpha$, and $\nu$. To fit the exponents, we use the standard procedure described in reference \protect\citeapp{clausetPowerLawDistributions2009}. First, we numerically find the maximum likelihood fits for a given distribution across a logarithmically-spaced range of lower cutoffs. For each lower cutoff, we calculate the Kolmogorov-Smirnov (KS) statistic (the maximum distance between the cumulative distribution functions) and choose the lower cutoff with the smallest statistic. This procedure defines how to determine the exponents and lower bounds from the distributions shown in Fig.~2.

To calculate significance, we sample from the power law fit. If there is a lower bound, we bootstrap sample from the data points below the lower cutoff to construct a full realization of a sample that is a combination of an unparameterized model below the cutoff and a power law above. We then run the same fitting procedure 2,500 times (again fitting the lower bound to each sample) to measure the distribution of the KS statistic. Thus, the KS statistic determines the $p$-value that we use for significance such that $p\geq0.1$ indicates that the observed distribution has a KS statistic smaller than 90\% of all bootstrapped samples, a strict test of significance \citeapp{clausetPowerLawDistributions2009}. Across much of the data, the distributions that we find satisfy this stringent criterion for significance demonstrating that the power law form is a convincing model for armed conflict statistics.

For the data that we consider in the main text where $b=140\,$km and $a=128$\,days, all the distributions (including for reports though it is not shown) except for extent are statistically indistinguishable from power laws with $p\geq0.1$. In general, it is not the case that every distribution for which we measure exponents satisfies this stringent criterion (Fig.~\ref{gr:p-values}). In the cases where the statistical test fails, often the power law model is a reasonable fit to the tail of the distribution. As a result, we can still measure an exponent though it may be a biased estimate. Such biases appear to be small because the estimated exponents across a range of spatiotemporal scales all take similar values (Fig.~2). Thus, across a large swathe of data, we find statistical evidence that power laws serve as accurate models when accounting for the spatiotemporal spread of conflict beyond individual events as have been investigated in other examples of armed conflict \citeapp{johnsonSimpleMathematical2013,picoliUniversalBursty2015,gillespieEstimatingNumber2017,clausetTrendsFluctuations2018}.

\begin{figure}[tb]\centering
	\includegraphics[width=.85\linewidth]{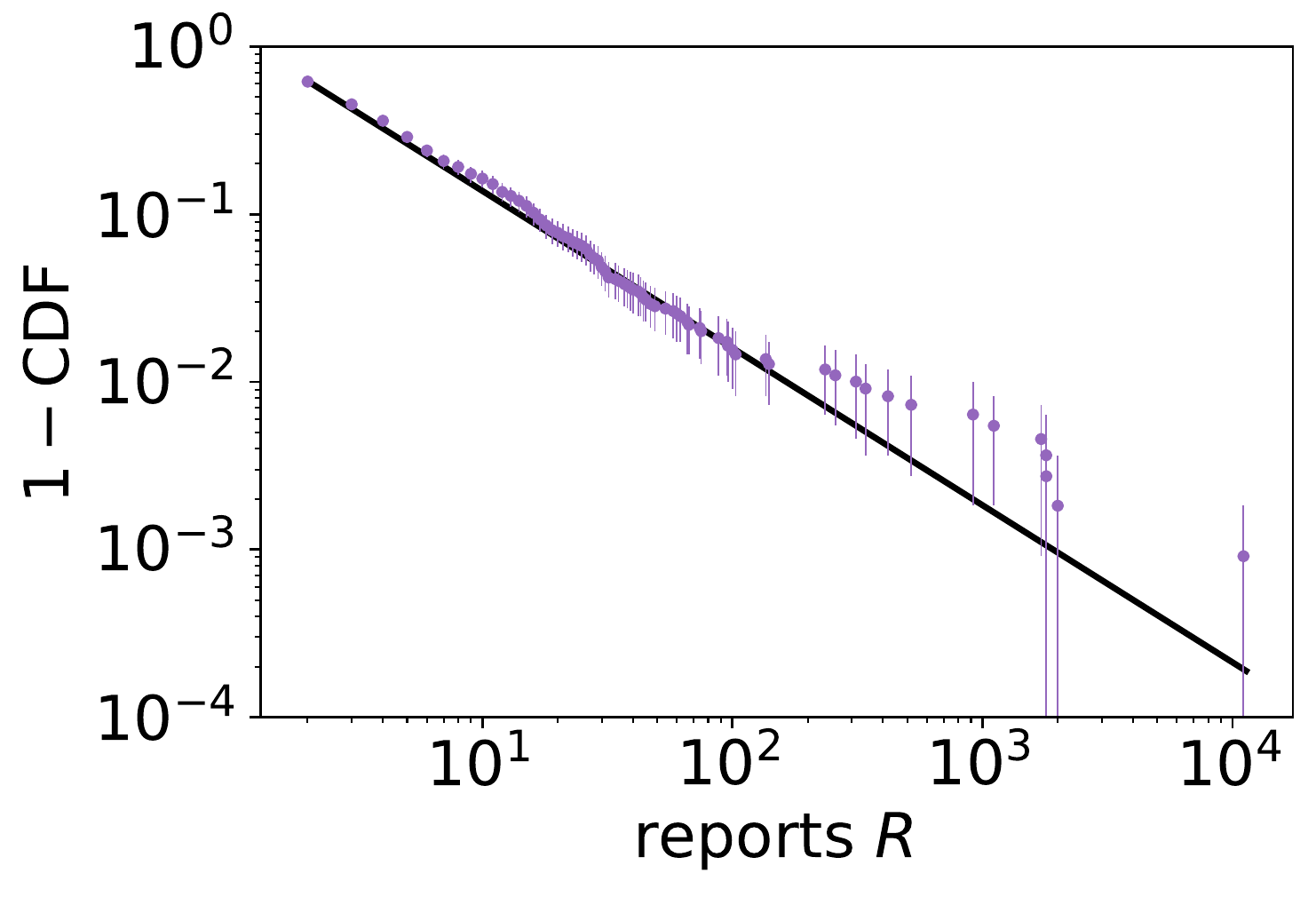}
	\caption{Distribution of reports (purple) compared with power law fit (black). Error bars represent 90\% bootstrapped confidence intervals. See Fig.~\ref{gr:fig1} for other conflict properties.}\label{gr:cdf reports}
\end{figure}

We measure the exponents for VAC and Riots/Protests and show them in Figs.~\ref{gr:vac exponents} and \ref{gr:riots exponents} for fixed $b=140\,$km and across the same range of $a$ as with Battles. 
In the top row of Figs.~\ref{gr:vac exponents} and \ref{gr:riots exponents}, the distributions for $a=128\,$days are all statistically indistinguishable from power laws except for $P(R)$ for Riots/Protests. Inspecting this distribution in more detail, we find a hump near the largest number of reports that deviates from the power law form.\footnote{Similar deviations from the power law are visible both for Battles and VAC reports distributions---though they are possibly statistical artifact such coincidence is noteworthy. Intriguingly, such humps are characteristic of finite-size effects in physical systems near the critical point where the largest avalanches ``pile up'' near the system size \protect\citeapp{sethnaCracklingNoise2001,cardyScalingRenormalization1996}. Although we do not do so here, it is tantalizing to consider what signals of (universal) finite-size corrections may appear in armed conflict data.} Thus, the evidence of strict adherence to a power law form is less clear for this particular distribution as we indicate with an asterisk in Table~\ref{tab:exponent table}. Nevertheless, we point out in Fig.~\ref{gr:riots exponents} that the exponents for the adjacent separation times $a=32\,$days and $a=256$\,days tightly bound the range of possible values for $a=128$\,days, which falls in between. Thus, we determine power law exponents for both VAC and Riots/Protests as we do with Battles using standard statistical tests and finding that these events are largely consistent with the power law hypothesis.

\section{Dynamical exponents}
Next, we measure the dynamical exponents by regression on the appropriate pair of scaling variables. A simple parameterization of the scaling relation is
\begin{align}
	X = a T^{\delta}
\end{align}
with coefficient parameter $a$ and exponent parameter $\delta$. If errors are multiplicative, the fitting procedure is equivalent to least-squares regression in logarithmic space. However, the typical regression problem only accounts for noise along the dependent variable (here $X$) which returns a solution that is not guaranteed to be symmetric about a fit to the inverse scaling relation
\begin{align}
	T = (X/a)^{1/\delta}.
\end{align}
This asymmetry presents ambiguity in the choice of which regression to use to measure the scaling exponents.

Instead, we define a fitting procedure that ensures symmetry about the inversion of the scaling relation. We minimize a symmetrized cost function that treats both $X$ and $T$ as dependent variables in turn
\begin{align}
	C(a,\delta, \sigma_{\rm X}, \sigma_T) &= \sum_{\rm i=1}^{\rm K} \left[\log X_{\rm i} - \delta\log T_{\rm i}- \log(a)\right]^2/\sigma_{\rm X}^2 + \notag\\
	&\left[ (\log X_{\rm i}-\log(a))/\delta - \log T_{\rm i}\right]^2/\sigma_T^2. \label{eq:log cost}
\end{align}
The variance parameters $\sigma_{\rm X}$ and $\sigma_T$ account for the possibility that magnitude of the noise along the $X$ dimension may be different than that of the noise along the $T$ dimension. By numerical simulation, we find that the regression procedure using the symmetrized cost function shows similar or less bias than the simple least-squares fit with noisy data, and thus we adopt Eq~\ref{eq:log cost} for fitting the dynamical exponents.

In Fig.~\ref{gr:fractal reg}, we show the results of regression using Eq~\ref{eq:log cost} to measure the dynamical scaling exponents for reports and fatalities of conflict avalanches. As we write in the main text, we measure $d_R/z = 2.0$, $d_F/z = 2.5$, and $1/z=0.8$ with the corresponding 90\% confidence intervals in Table~\ref{tab:exponent table}.

\begin{figure}[tb]\centering
	\includegraphics[width=.9\linewidth]{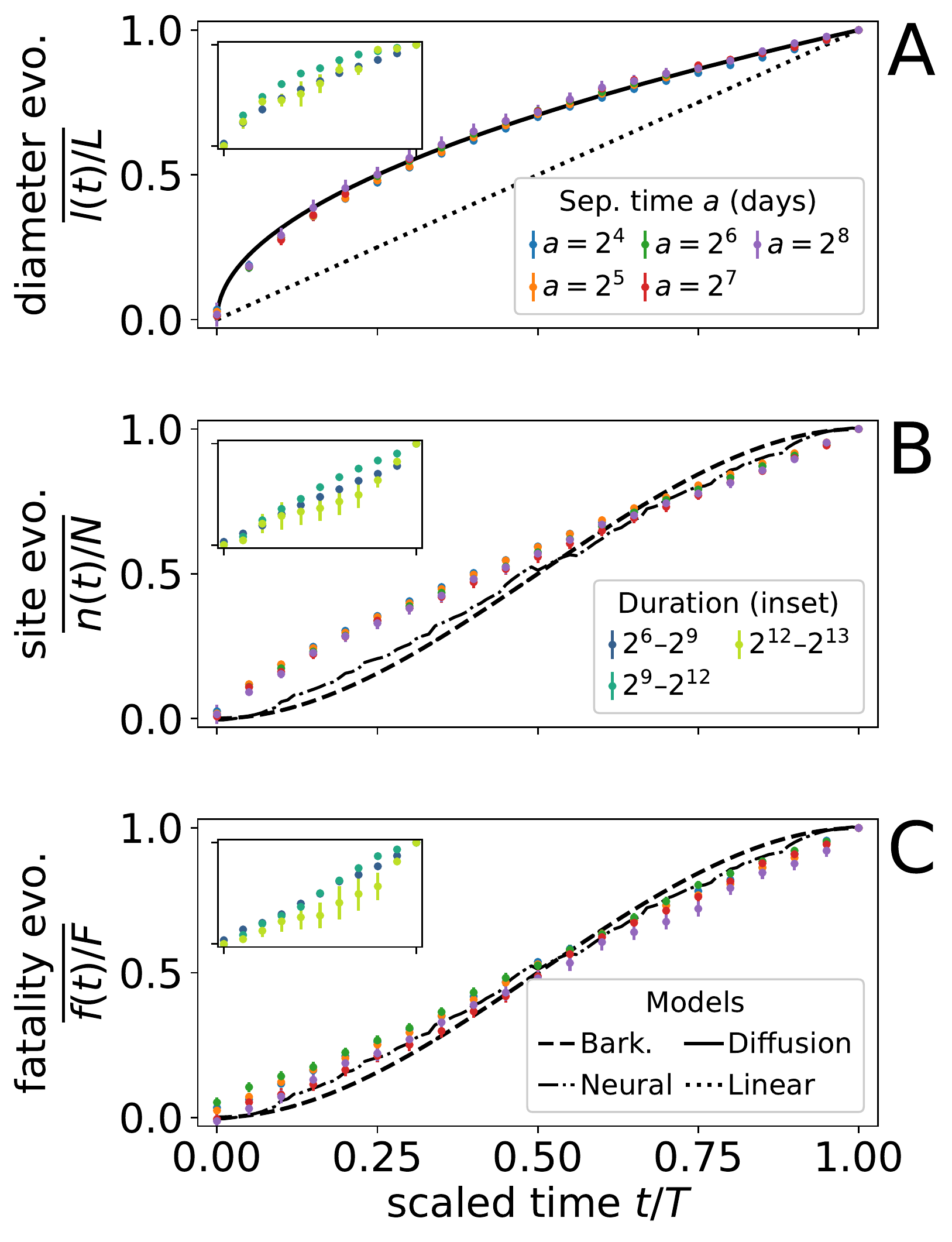}
	\caption{Temporal profiles for VAC events converge to a universal profile similar to that of Battles from Fig.~3.}\label{gr:vac dynamics}
\end{figure}

\begin{figure}[t!]
	\includegraphics[width=.9\linewidth]{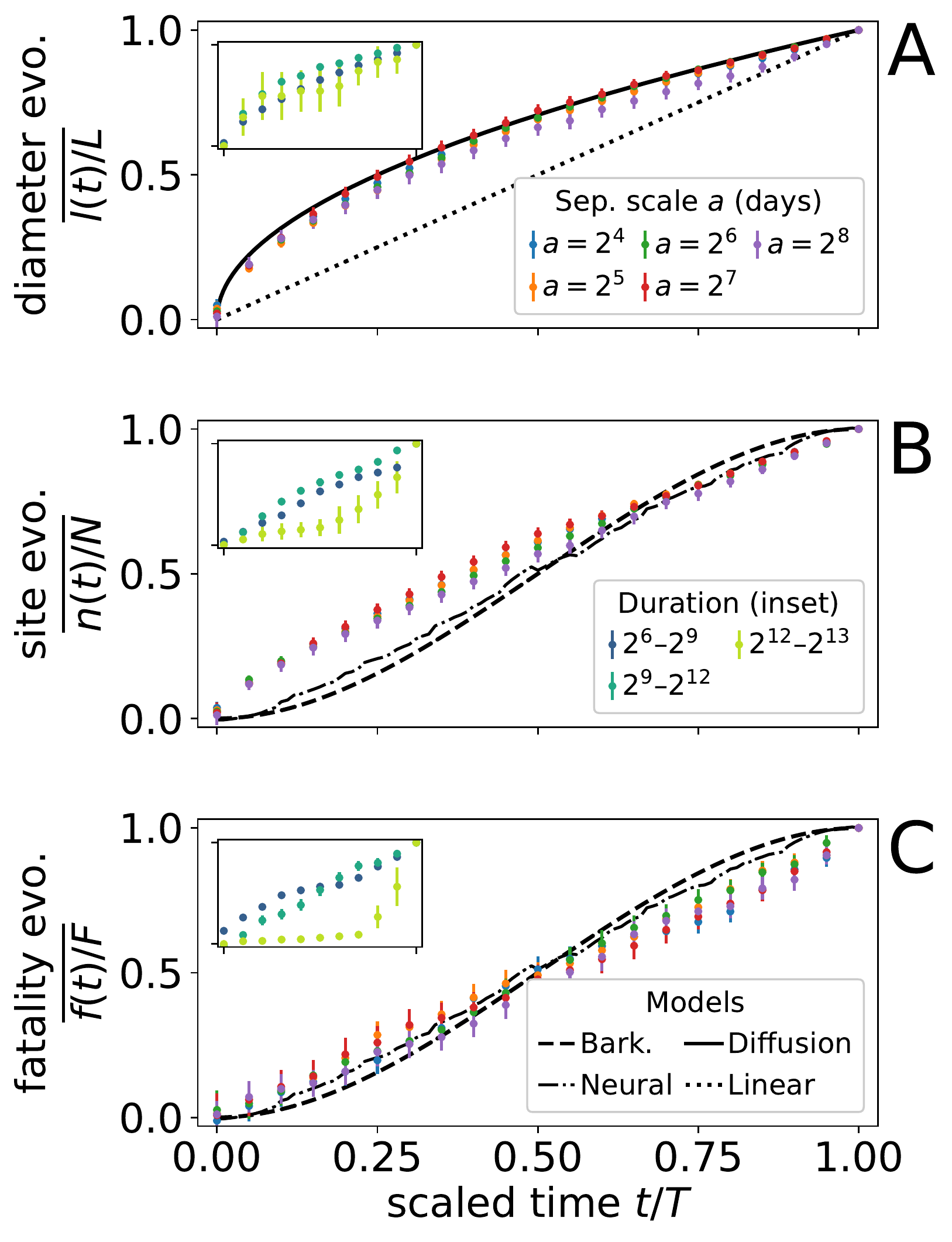}
	\caption{Temporal profiles for Riots/Protests. (inset in A) The profile for the longest conflict avalanches with durations $2^{11}$--$2^{13}$\,days show significant deviations away from the universal profile. (B) Riots/Protests typically have a much smaller fatality count and thus a small sample of temporal profiles. As a result, the profiles are not as well estimated as those for Battles and VAC and show large variation.}\label{gr:riots dynamics}
\end{figure}

\section{A cumulative temporal profile}\label{sec:cum profile}
We take a non-parametric approach to showing the collapse of conflict temporal profiles using a cumulative curve because of our small data set. In contrast, rate profile curves are often shown elsewhere such as with neural avalanches \citeapp{papanikolaouUniversalityPower2011,friedmanUniversalCritical2012,timmeCriticalityMaximizes2016,poncealvarezWholeBrainNeuronal2018}. Whereas controlled experiments permit observation of multiple systems with $\gtrsim10^4$ avalanches, we have at most $K<10^3$ avalanches above the lower cutoffs of $T\geq8$\,days, $R>2$, and $F>2$. For large $b$ and $a$, we have even fewer avalanches $K<10^2$. For the temporal bins shown in Fig.~3, the number of samples ranges from $<10$ to a few hundred in the best sampled bins even with logarithmic spacing. As a result, the rate temporal profiles show considerable variability, and the statistical similarity between the profiles is overshadowed by visual noise.

\begin{figure}[tb]
	\includegraphics[width=.8\linewidth]{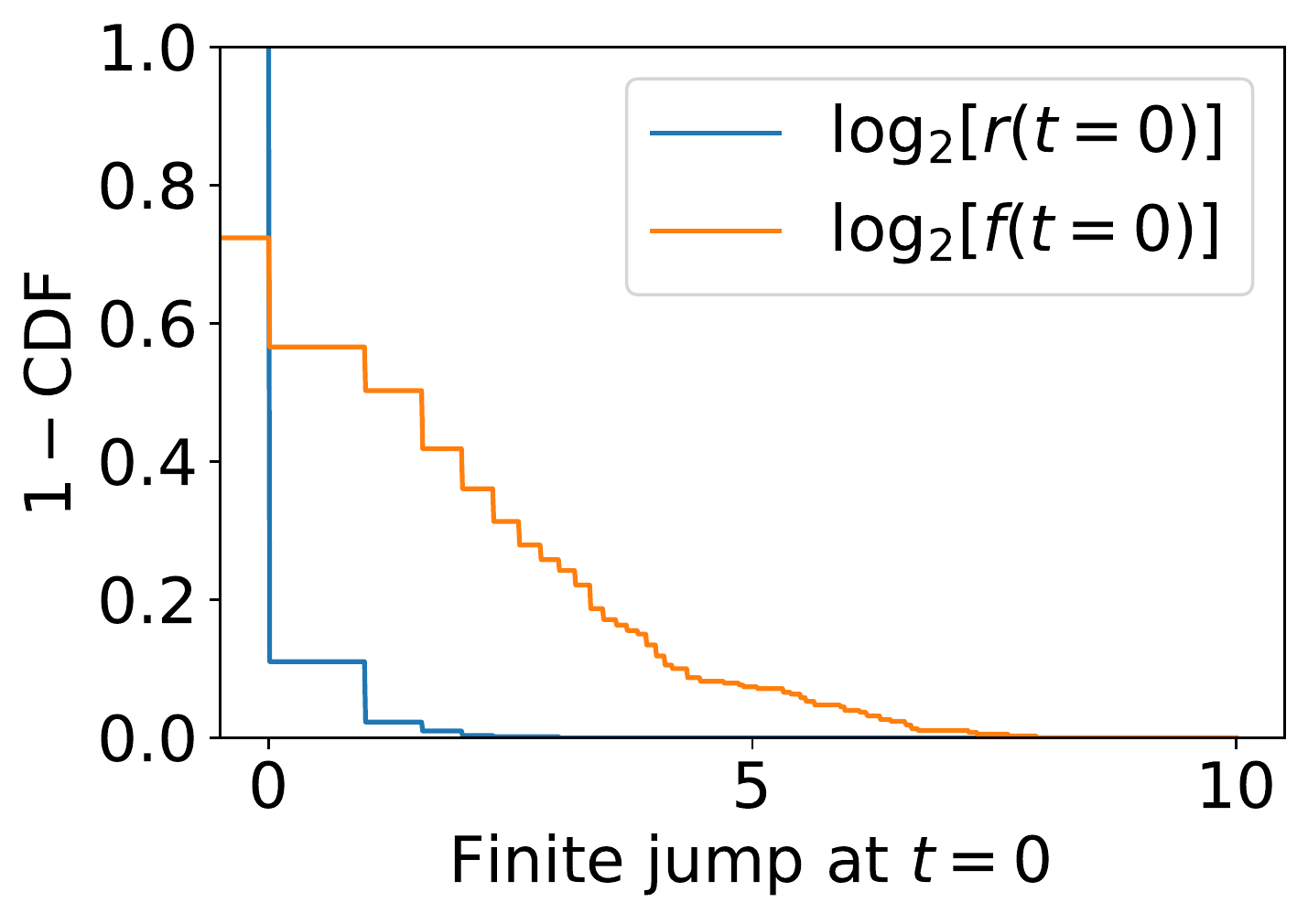}
	\caption{Distribution of finite jump at $t=0$ in temporal profiles. For report profiles (blue), nearly all conflicts only involve a single report on the first day which can be accounted for as a lattice bias that decay as $1/R$. Fatalities (orange) show a much wider distribution since it possible for any number of fatalities to occur on the first day. The average $\br{1/f(t=0)}$ can again be accounted for by a lattice bias that depends only the number of reports $\br{1/R}$.}\label{gr:finite jumps}
\end{figure}

To construct the cumulative profiles, we use the right-handed cumulative distribution function counting the number of events scaled by total size of the conflict (either by $R$ or $F$) and by the total duration $T$. By definition, all the report profiles $\int_0^t r(t')\,dt'$ must end at 1 and they must start at $1/R$ (Fig.~\ref{gr:finite jumps}). This offset constitutes a {\it lattice} bias that disappears geometrically as $R\rightarrow \infty$, but many of our profiles involve small avalanches. To account for this bias, we subtract from the profile the value $1/R$, again subtract $1/R$ from $t=T$, and then scale the profile such that it ends at 1. As a result, profiles of avalanches with reports $R=2$ are meaningless and thus are excluded from this analysis. The same lattice bias appears in rate profiles since avalanches by definition start with at least one event per time bin. As with the cumulative profile, the finite jump decays geometrically to 0 with the size of the avalanche. Although we are not aware of any explicit mention of such corrections with neural avalanches---they are typically left uncorrected in collapsed profiles perhaps because neural avalanches are much bigger---the prevalence of small conflict avalanches means that accounting for such biases is essential for capturing the temporal profile collapse for sizes.

For fatalities, however, subtracting such bias per avalanche is an ill-posed solution because some reports include no fatalities leading to the possibility of negative cumulative fractions. Indeed, any number of fatalities may occur at $t=0$ so there is no {\it a priori} reason to account for a lattice effect of $1/F$ (Fig.~\ref{gr:finite jumps}). Yet, we find a substantial fraction of events occur on the first day, accounting for about 30\% of all fatalities for conflicts of duration $T\leq a$ and 10--20\% in conflicts $T>a$ and decreasing in a roughly geometric manner with conflict duration.
Motivated by the nearly linear profile between the endpoints, we look over all fatality profiles and start with the assumption that fatalities occurred with uniform probability across all $R$ reports filed during a conflict avalanche. In other words, such a null model would imply that an average fraction of $\br{1/R}$ fatalities on the first and last days of a conflict avalanche of duration $T$. Similar to report profiles, we find that the sizable jumps at $t=0$ and $t=T$ can be almost completely accounted for by an analogous $\br{1/R}$ lattice bias. Thus, we find a collapse of the temporal profiles for both reports and fatalities after accounting for lattice bias incurred by the discrete nature of conflicts in the data.

\begin{figure}[tb]
	\includegraphics[width=.95\linewidth]{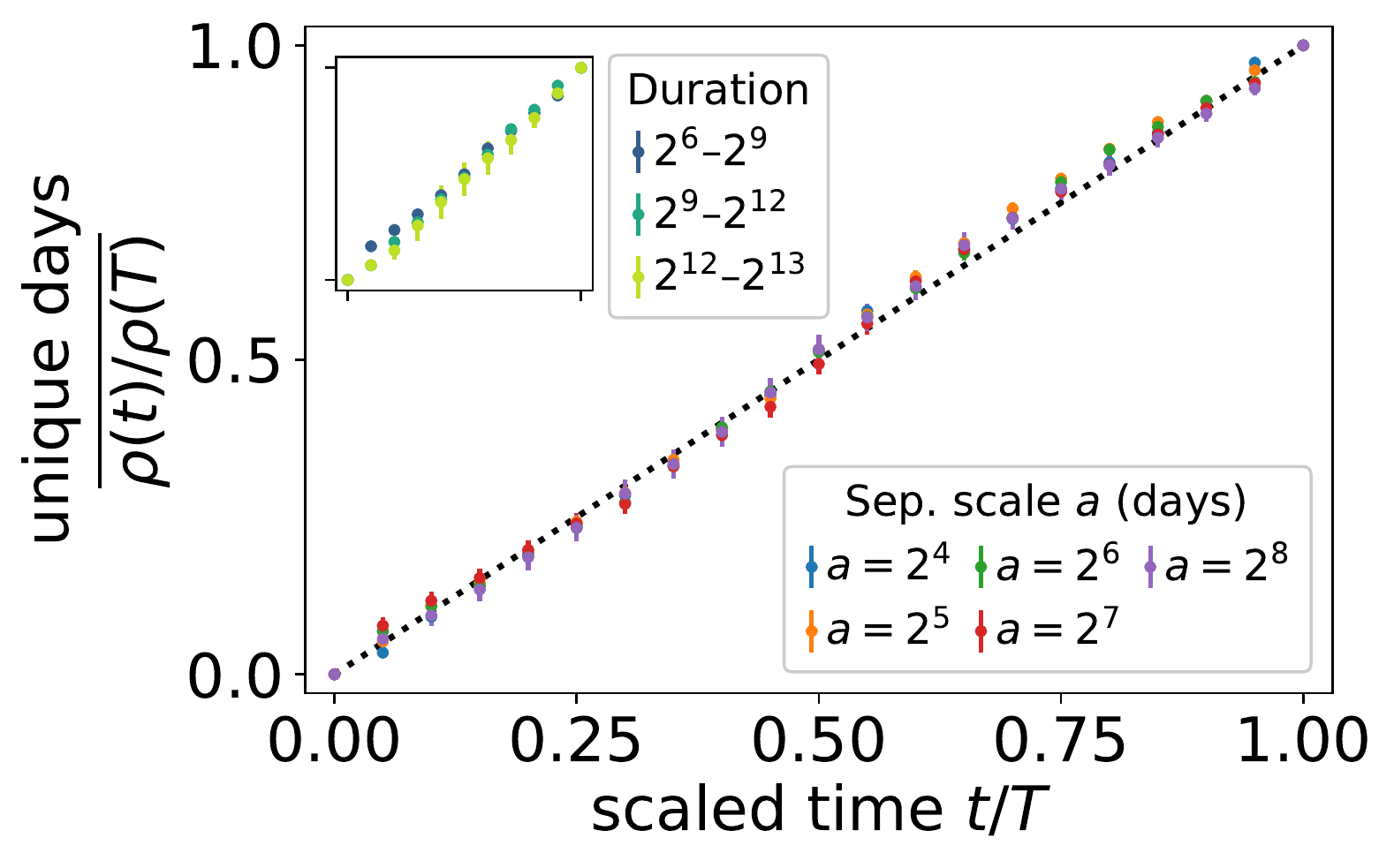}
	\caption{Battle conflict avalanche temporal profiles counting the cumulative number of unique days on which events have occurred $\int_0^t \br{\rho(t')/P}\,dt'$.}\label{gr:profile by day}
\end{figure}

We find that the temporal profiles for VAC and Riots/Protests resemble those of Battles as pictured in Fig.~\ref{gr:vac dynamics} and \ref{gr:riots dynamics}. Although the smaller size of Riots/Protests events introduces more variability, we find most of the profiles are largely consistent with those of Battles: the temporal profiles for long avalanches are nearly linear and the geographic extent grows like diffusion. One notable outlier is the profile of the longest conflict avalanches for Riots/Protests that seem to accelerate near the ends of the profiles---though the reasons for such divergence are unclear. Overall, this coincidence in temporal profiles leads to the surprising possibility that the dynamics of armed conflict are largely analogous across multiple kinds of conflict when observed over sufficiently large scales.

\begin{figure}[tb]
	\includegraphics[width=\linewidth]{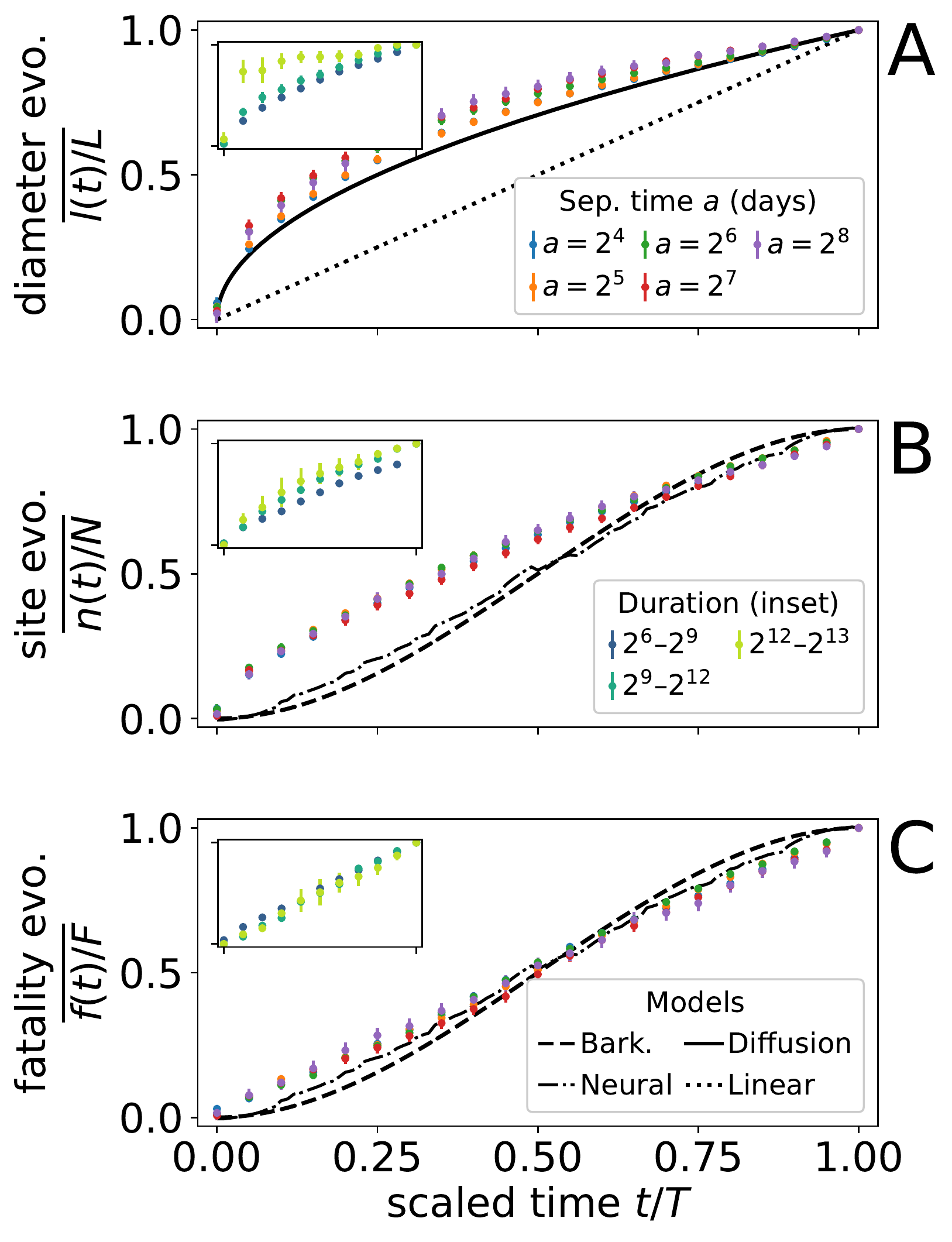}
	\caption{Battles temporal profiles after time shuffle. Fatality evolution remains largely unchanged from Fig.~3. Reports evolution remains largely unchanged (not shown).}\label{gr:profile shuffled}
\end{figure}

\section{Temporal profile reflects conflict rate}

We find that the shape of conflict avalanche profiles can be traced back to the rate at which events are observed. In Fig.~\ref{gr:profile by day}, we show the temporal profiles of the cumulative number of unique days on which events happen during a conflict avalanche for $b=140\,$km across the range of $16\leq a\leq 256$\,days---equivalent to setting every event to reports $R=1$ and accounting for the lattice correction discussed in Section~E. We denote this unique day rate profile as $\rho(t)$. The fact that the unique day rate profiles resemble that of Battles suggests that the report profile is dominated by the timing between reports that are filed rather than the number of reports on any given day.

As a further check of this hypothesis, we time shuffle the reports and fatalities within each conflict avalanche while keeping fixed the days on which events were reported (i.e., events can only occur on a day when at least one report was previously filed after the shuffle). If it is the case that events of different reports were preferentially clustered at certain points during a conflict avalanche, such a shuffling procedure should flatten the profiles. Unsurprisingly, we show in Fig.~\ref{gr:profile shuffled} that the nearly linear profiles remains largely unchanged for fatalities (it is also the case for reports). In contrast, geographic spread is clearly changed, reaching the boundaries faster and indicating that avalanches preferentially spread out from some epicenter. These results are yet again consistent with the hypothesis that report and fatality profiles are dominated by the rate of events, distinct from the geographic spread of conflicts.

\section{Activated random walkers, percolation growth, and other models}\label{sec:other models}
\begin{figure}[tb]\centering
	\includegraphics[width=\linewidth]{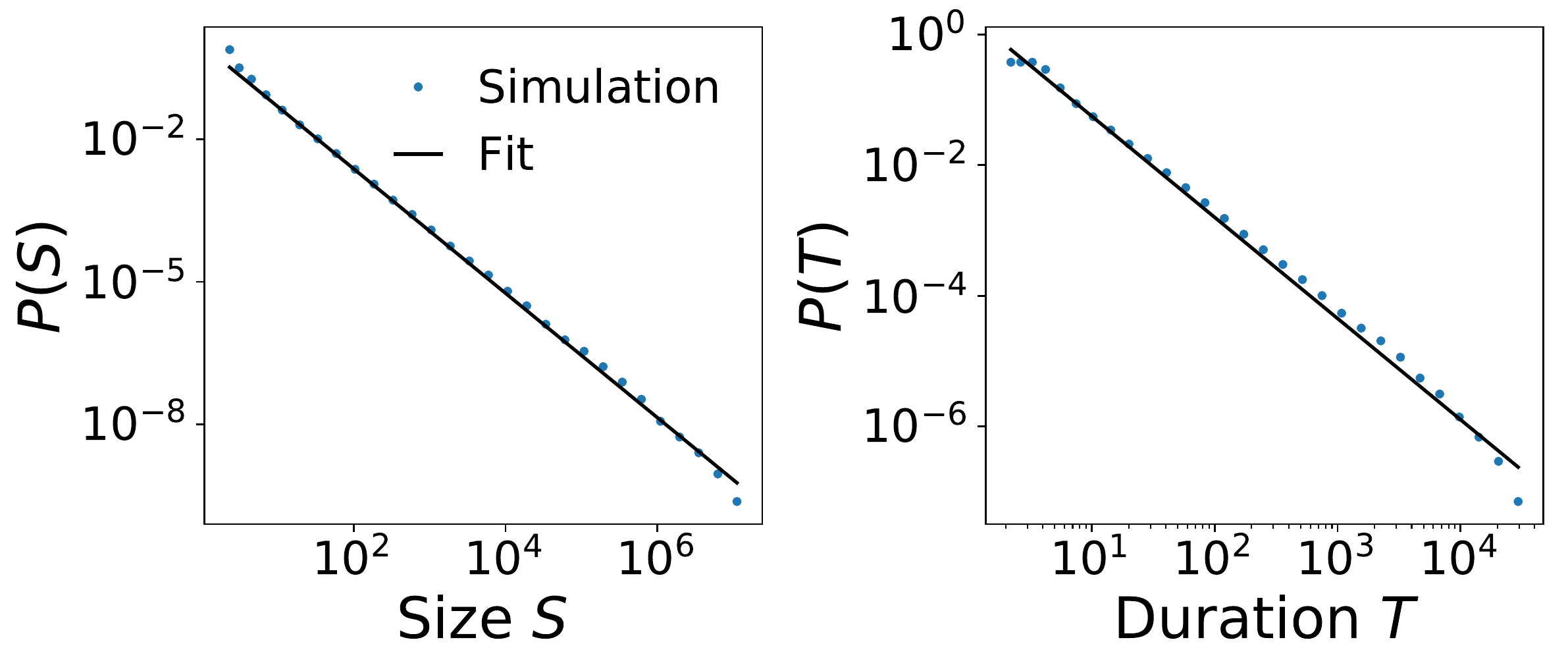}
	\caption{Distributions of size and duration for activated random walkers model in 2D \protect\citeapp{dickmanPathsSelforganized2000}.}\label{gr:arw sim}
\end{figure}

We simulate the activated random walkers (ARW) model described in reference \citeapp{dickmanPathsSelforganized2000}. The model consists of ``walkers,'' or particles, living on lattice sites that are inactive when alone but are activated when there are multiple walkers on the same site. At every site with multiple walkers, two walkers move to randomly chosen neighbors. As long as any walkers are active, the cascade continues and grows in size $S$, measured by the cumulative number of walkers that move at each step, and duration $T$, measured by the number of simultaneous updates over the entire lattice. To produce the distributions we show in Fig.~\ref{gr:arw sim}, we used a square lattice with edge length $l=10^3$ with free boundary conditions such that walkers that exceed the boundaries disappear. Whenever the dynamics stop, we add a walker at a random site which may or may not start the dynamics again. Using maximum likelihood, we find the distribution exponents for size $\tau = 1.31$ with lower cutoff of 60 and duration $\alpha= 1.55$ with lower cutoff of 45 over $10^4$ samples.

``Percolation growth'' in Table~1 refers to the growth of a percolation cluster on a 2D square lattice \citeapp{staufferIntroductionPercolation1994,clarScalingLaws1994}. 
Though percolation is a static, geometric transition and not dynamical, avalanche patterns can share properties with percolation (such as forest fires; references \citeapp{staufferIntroductionPercolation1994} and \citeapp{clarScalingLaws1994}). The model we refer to is akin to the way that forest fires in the model grow on connected clusters of trees at the critical point. After seeding a lattice with an occupied site at the origin, we grow a percolation cluster by occupying the neighbors of any occupied site with some probability $p$ that the connecting bond is ``open.'' We count time in units of shells such that the unoccupied neighbors of any of the currently occupied sites are simultaneously occupied in one time step. Using a square lattice with edge length $l=10^4$, open bond probability $p=0.49$, $K=10^3$\,samples, and fitting to trajectories with duration $T\geq10$, we recover the exponents in reference \citeapp{clarScalingLaws1994}.

Besides these two models, we show exponents measured from a variety of other models and experimental systems that show cascade dynamics in Table~\ref{tab:exponent table}. Surprisingly, one model aligned across multiple exponent values is percolation growth. Despite this alignment, percolation growth, like the other physical models, is an incomplete analogy for the multi-faceted aspect of armed conflict. For example, there is ambiguity between how cascade size should be compared with the multiple measures of armed conflict size.\footnote{In this sense, armed conflict is more akin to neural avalanches, where a single neuron may contribute multiple times to a cascade, though again the exponents disagree with what we measure.} As we argue in the main text, armed conflict consists of two separate processes for geographic spread and social growth, separation that is supported by measured exponents. Thus, physical cascade models may provide an apt analogy for some subset of armed conflict features (e.g., the distribution of fatalities with exponent tantalizingly close to $\tau=3/2$) but are not alone sufficient to capture the multiple properties of conflict avalanches we explore.

Going beyond alignment with exponents, the temporal profiles hint at the underlying dynamics generating conflict avalanches. For comparison, we show profiles of canonical systems with self-similar avalanches like Barkhausen noise and an example of a neural culture in Fig.~\ref{gr:fig3}. These tend to accelerate in the middle whereas average size and fatality profiles for conflict avalanches tend to evolve at a more linear pace. Flat profiles can indicate dissipative effects that suppress large events as with demagnetizing fields in Barkhausen noise \citeapp{papanikolaouUniversalityPower2011}. Yet, flattening is also a feature of both subcritical and supercritical cascades that spontaneously end---though such profiles will fail to collapse \citeapp{hindesEpidemicExtinction2016,gleesonTemporalProfiles2017}. Thus, the mapping between dynamics and profile is many-to-one, but we can rule out analogues of properties that, for example, generate asymmetric profiles such as eddy currents in magnetic materials \citeapp{papanikolaouUniversalityPower2011}, certain networks like in disassociated neural cultures \citeapp{friedmanUniversalCritical2012}, or variations in birth-death processes \citeapp{gleesonTemporalProfiles2017}.
In contrast, we find that spatial extent grows in a strongly nonlinear and asymmetric fashion as shown in Fig.~\ref{gr:fig3}C. This profile is closely described by the average linear extent of a convex hull of planar Brownian walkers \citeapp{kotDispersalData1996,randon-furlingConvexHull2009}, perhaps related to properties of generalized diffusion models used to describe other conflict data sets \citeapp{zammit-mangionPointProcess2012}.
More generally, these profiles are compatible with Markovian cascades on networks indicating that such dynamics may come to dominate in long conflict avalanches.

A fascinating and debated question that arises when comparing armed conflict statistics with physical models is whether or not conflict avalanches sit near a critical point \citeapp{robertsFractalitySelfOrganized1998}. Indeed the notion that armed conflict might represent a kind of self-organized criticality has intuitive appeal in the sense that social tension might drive social systems to a point where conflict is likely to break out \citeapp{cedermanModelingSize2003}. As we show in the main text, however, we find evidence that scaling may arise from geography and that some of the scaling we find is the result of strong, global correlations in conflict intensity. This does not rule out the possibility of ``criticality'' in conflict avalanches, but suggests that multiple related aspects may contribute that may result from different mechanisms. We leave deeper exploration of this connection for later work.

\bibliographystyleapp{unsrt}
\bibliographyapp{si}  


\end{document}